\documentclass[11pt]{article}
\usepackage{amsmath,latexsym,amssymb,epsfig}
\usepackage{graphicx}

\def\ltsim{\lower3pt\hbox{$\, \buildrel < \over \sim \, $}} 
\def\gtsim{\lower3pt\hbox{$\, \buildrel > \over \sim \, $}} 
\newcommand{\be}{\begin{equation}} 
\newcommand{\ee}{\end{equation}} 
\def\ga{\mathrel{\raise.3ex\hbox{$>$\kern-.75em\lower1ex\hbox{$\sim$}}}} 
\def\la{\mathrel{\raise.3ex\hbox{$<$\kern-.75em\lower1ex\hbox{$\sim$}}}}

\makeatletter 
\@addtoreset{equation}{section} 
\makeatother

\def\mneut{m_{\tilde{\chi}^0_1}}

\def\chargm{\tilde{\chi}^-_1}

\def\neu{\tilde{\chi}^0}
\def\neut{\tilde{\chi}^0_1}
\def\charg{\tilde{\chi}^+_1}
\def\neutt{\tilde{\chi}^0_2}
\def\sip{\sigma^{SI}_{\chi p}}
\def\chargt{\tilde{\chi}^+_2}
\def\binop{\tilde{B'}}
\def\winop{\tilde{W'}}
\def\bino{\tilde{B}}
\def\wino{\tilde{W}}

\begin{document}
\baselineskip=16pt

\begin{titlepage}

\vskip 2cm
\begin{flushright}
 LAPTH-1327/09
\end{flushright}
\begin{center} 
\vspace{0.5cm}
\Large {\sc Dark Matter with Dirac and Majorana Gaugino Masses}
\vspace*{5mm} 
\normalsize

{\bf
G. Belanger$^*$
\footnote{belanger@lapp.in2p3.fr},
K.~Benakli$^\dagger$ 
\footnote{kbenakli@lpthe.jussieu.fr},
M.~Goodsell$^\dagger$ 
\footnote{goodsell@lpthe.jussieu.fr},
C.~Moura$^\dagger$
\footnote{moura@lpthe.jussieu.fr}
and
A. Pukhov$^\ddagger$
\footnote{pukhov@lapp.in2p3.fr}
}

\smallskip 
\medskip 
\it{$^*$ LAPTH, Univ. de Savoie, CNRS, B.P. 110, F-74941 Annecy-le-Vieux, France} \\
\it{$^\dagger $ LPTHE, Universit\'e Pierre et Marie Curie - Paris VI, France}\\
\it{$^\ddagger$SINP, Moscow State University, Moscow 119992, Russia}

\vskip0.6in 
\end{center}

\begin{abstract}
We consider the minimal supersymmetric extension of the  Standard Model allowing both Dirac and Majorana gauginos. The Dirac masses are obtained by pairing up  extra chiral multiplets: a singlet $\mathbf{S}$ for $U(1)_Y$, a triplet $\mathbf{T}$ for $SU(2)_w$ and an octet $\mathbf{O_g}$ for $SU(3)_c$ with the respective gauginos. The  electroweak symmetry breaking sector is modified by the couplings of the new fields $\mathbf{S}$ and $\mathbf{T}$ to the Higgs doublets. We discuss two limits: i) both the adjoint scalars are decoupled with the main effect being the modification of the Higgs quartic coupling; ii) the singlet remaining light, and due to its direct coupling to  sfermions, providing a new contribution to the soft masses and inducing new decay/production channels. We discuss the LSP in this scenario; after mentioning the possibility that it may be a Dirac gravitino, we focus on the case where it is identified with the lightest neutralino, and exhibit particular values of the parameter space where the 
relic density  is in agreement with WMAP data. This is illustrated for different scenarios where the LSP is  either a bino (in which case it can be a Dirac fermion) or bino-higgsino/wino mixtures. We also point out in each case the peculiarity  of the model with respect to dark matter  detection experiments. 
\end{abstract}

\end{titlepage}

\section{Introduction}
\setcounter{footnote}{0}

A remarkable fact of nature is that the light fundamental fermions appear in the smallest representations, the singlet or fundamental, of the Standard Model $SU(3)_c\times SU(2)_w\times U(1)_Y$  symmetry group. Could larger representations be present at higher energies and be discovered at the LHC? Such particles are in fact predicted by low energy supersymmetry as  gauginos, superpartners of the gauge vector bosons. These gauginos can appear as fermions of Majorana (with only two degrees of freedom)  or Dirac (with four degrees of freedom)  type. In the Minimal Supersymmetric extension of the Standard Model (MSSM)  the gauginos are Majorana. Obtaining Dirac masses would require pairing them with additional Dirac Gaugino adjoint (henceforth DG-adjoint) states: a singlet $\mathbf{S}$ for $U(1)_Y$, a triplet $\mathbf{T}$ for $SU(2)_w$ and an octet $\mathbf{O_g}$ for $SU(3)_c$.

Construction of models with spontaneous  breaking of supersymmetry may lead to  non-minimal extensions of the Standard Model  that include Dirac gauginos. The presence of the required DG-adjoints in the light spectrum is often motivated by the presence of an underlying $N=2$ supersymmetry, that pairs them with the vector multiplets. 
Such a scenario was  first introduced by Fayet \cite{Fayet:1978qc} as a way to give masses for gluinos while preserving R-symmetry\footnote{ In \cite{Fayet:1978qc} $R$-symmetry was later broken by Majorana masses for the DG-adjoints in order to avoid tachyonic masses for their scalar components. An issue that was recently solved in \cite{Benakli:2008pg} and independently for an explicit model in \cite{Amigo:2008rc}.}. The soft nature of these masses requires a modification of the interaction as was shown using   $D$-term breaking  by \cite{Polchinski:1982an, Gates:1983nr}. More recently,  Dirac gauginos have arisen in models with an extra dimension where supersymmetry is broken by a Scherk-Schwarz mechanism (see for example \cite{Antoniadis:1992fh, Pomarol:1998sd}).  More precisely, they are a combination of two Majorana fermions with mass  given by $1/2R$ (half of the compactification scale $1/R$), one given by the (mass shifted)  massless mode and the other from the (mass shifted) first Kaluza-Klein state, thus,  the DG-adjoints originate 
as ``half of the first Kaluza-Klein excitation".  It was noted there that the soft masses are UV finite and do not exhibit the usual logarithmic sensitivity to the UV cut-off \cite{Antoniadis:1998sd}. This property was shown to persist in four dimensional models as being peculiar to the Dirac nature of the gaugino masses, and denoted as supersoft in \cite{Fox:2002bu,Nelson:2002ca, Chacko:2004mi,Carpenter:2005tz,Nomura:2005rj} where the important phenomenological implications of $D$-term supersymmetry breaking were first outlined.  They  were further studied in  constructions  of non-supersymmetric intersecting brane models \cite{Antoniadis:2005em, Antoniadis:2006eb, Antoniadis:2006uj}. More recent  examples arise from the possibility of using calculable $R$-symmetric  models \cite{Gates:1983nr, Hall:1990hq, Kribs:2007ac, Amigo:2008rc, Marques:2009yu, Blechman:2009if}.  For instance, it was pointed out in \cite{Kribs:2007ac} that they lessen the flavor problem in supersymmetric theories. This  renewal of interest in
  such models has been  motivated by the work of ISS \cite{Intriligator:2006dd}. Furthermore,  deforming  the  ISS model with explicit breaking of $R$-symmetry could  leave states in adjoint representations \cite{Zur:2008zg} and allow the simultaneous presence of both Majorana and Dirac masses for the gauginos. The generation of Dirac gaugino masses can also be included \cite{Benakli:2008pg} in the framework of ``general gauge mediation" \cite{Meade:2008wd,Carpenter:2008wi,Ooguri:2008ez,Distler:2008bt,Intriligator:2008fr,Buican:2008ws}.

In this work, we will study a minimal extension of the MSSM that incorporates both   Majorana and Dirac gaugino masses. The field content is that of the MSSM, supplemented with the DG-adjoints. The MSSM renormalisable Lagrangian is then  supplemented by i) the DG-adjoint kinetic and mass terms, ii) the Dirac gaugino masses, iii) coupling of the singlet $\mathbf{S}$ and the triplet $\mathbf{T}$ to the Higgs doublets with strength $\lambda_S$ and $\lambda_T$ respectively, iv) the DG-adjoint scalar soft masses and trilinear terms.

For phenomenological issues the strength of the coupling of the Higgs to the DG-adjoint is of  particular importance. In general such couplings are arbitrary and subject to diverse phenomenological bounds as discussed in \cite{Nelson:2002ca}. Inspired by extra dimensional models, and in order to make the role of $N=2$ manifest, one can assume that the two MSSM Higgs doublets originate from a single hypermultiplet of an underlying $N=2$ supersymmetry. These models with combined $N=2$ and $N=1$ sectors   were introduced in  \cite{ Antoniadis:2006uj,:2008gva}.  In this case, the couplings $\lambda_S$ and $\lambda_T$ are related to the gauge couplings by $N=2$ supersymmetry. In this work, we will arbitrarily take the values of both couplings to go from zero to their tree level $N=2$ value to illustrate the model.

Some particular signatures of these models at collider experiments have been studied in 
\cite{Choi:2008pi,Choi:2008hh,Choi:2008ub,Kramer:2009kp, Nojiri:2007jm, Plehn:2008ae}. They  stressed that the Dirac/Majorana nature of the gluinos affects the distribution of produced squark states. It was pointed out in  \cite{Choi:2008ub} that the pair creation of scalar octets at the LHC will have a peculiar signature through cascade decays giving rise to a burst of eight or more jets together with four LSPs as well as through a resonance due to the  decays into gluons or a $t \bar t$ pair at the one-loop level.

In this work we will focus mainly on the fate of dark matter in this class of models. It is by now an important quality of the $R$-parity preserving versions of the MSSM that they provide a natural candidate for Dark Matter, the lightest supersymmetric particle  (LSP). A particular case is when the LSP is identified with the lightest neutralino. While in the MSSM the latter is a linear combination of the four neutral fermions, given by the bino, the wino and the two Higgsinos, it is now a linear combination of six states, the singlet and triplet fermions adding to the previous four. In this work we shall not try to give an exhaustive discussion of such a situation but try to answer such questions as: Are there parameter regions where the neutralino is a good dark matter candidate? How does the situation compare to the MSSM? An early study \cite{Hsieh:2007wq} of dark matter in a related model to that considered here focussed on the bino LSP, for a very particular case with
vanishing $\lambda_S$ and $\lambda_T$, and assumed dominance of the exchange of sfermions and thus neglecting, for example, the exchange of Higgs or gauge bosons. It concluded that the bino annihilation cross section can be enhanced which might help to obtain a smaller relic abundance than in the MSSM.

The present paper is organized as follows: Section 2 presents the model and defines our notations and conventions. Section 3 gives a comprehensive discussion, which we believe to be missing in the present literature,  of the electroweak breaking sector. The discussion follows the same lines as in \cite{Antoniadis:2006uj} which specializes to a  model with combined  $N=2$ and $N=1$ sectors in the limit of very large soft masses for the scalar components of $\mathbf{T}$ and $\mathbf{S}$. It differs from the usual extension of the MSSM by the couplings to the Higgs of a singlet, whose vacuum expectation value (vev) is not necessarly related to the supersymmetric Higgs mass term $\mu$,
and  a triplet (see \cite{Espinosa:1991wt,Espinosa:1991gr,FelixBeltran:2002tb,BarradasGuevara:2004qi,DiazCruz:2007tf,DiChiara:2008rg}). While the triplet scalar is required to be heavy by electroweak precision tests, the singlet can be either very heavy and integrated out or remain light with sensible mixing with the ordinary MSSM Higgs  states. We discuss both limits. In section 5, we briefly discuss the gravitino LSP and then focus on the case of a neutralino LSP. The corresponding mass matrix is exhibited and the nature of the lightest eiganstates is studied for some particular limits, in particular the necessary condition for having a Dirac fermion LSP is given. Section 6 presents numerical results for the relic adundance and the corresponding signature at direct/indirect detection experiments are stressed.

\section{The model}

\begin{table}[htb]
\begin{center}
\begin{tabular}{c|c|c|c|c|c}
\hline
Names  &                 & Spin 0                  & Spin 1/2 & Spin 1 & $SU(3)$, $SU(2)$, $U(1)_Y$ \\ 
\hline
&  &   &      &  &   \\ 
Quarks  & $\mathbf{Q}$   & $\tilde{Q}=(\tilde{u}_L,\tilde{d}_L)$  & $(u_L,d_L)$ & & (\textbf{3}, \textbf{2}, 1/6) \\ 
   & $\mathbf{u^c}$ & $\tilde{u}^c_L$              & $u^c_L$     & & ($\overline{\textbf{3}}$, \textbf{1}, -2/3) \\ 
($\times 3$ families) & $\mathbf{d^c}$ & $\tilde{d}^c_L$     & $u^c_L$     & & ($\overline{\textbf{3}}$, \textbf{1}, 1/3)  \\ 
\hline
Leptons & $\mathbf{L}$ & ($\tilde{\nu}_{eL}$,$\tilde{e}_L$) & $(\nu_{eL},e_L)$ & & (\textbf{1}, \textbf{2}, -1/2) \\ 
($\times 3$ families) & $\mathbf{e^c}$ & $\tilde{e}^c_L$    & $e^c_L$          & & (\textbf{1}, \textbf{1}, 1)  \\ 
\hline
Higgs & $\mathbf{H_u}$ & $(H_u^+ , H_u^0)$ & $(\tilde{H}_u^+ , \tilde{H}_u^0)$ & & (\textbf{1}, \textbf{2}, 1/2)  \\ 
  & $\mathbf{H_d}$ & $(H_d^0 , H_d^-)$ & $(\tilde{H}_d^0 , \tilde{H}_d^-)$ & & (\textbf{1}, \textbf{2}, -1/2) \\
\hline
Gluons & $\mathbf{W_{3\alpha}}$ & & $\lambda_{3\alpha} $                       & $g$              & (\textbf{8}, \textbf{1}, 0) \\ 
&   & & $  [\equiv \tilde{g}_{\alpha}]$                       &                &  \\ 
&  &   &      &  &   \\ 
W    & $\mathbf{W_{2\alpha}}$ & & $\lambda_{2\alpha} $ & $W^{\pm} , W^0$  & (\textbf{1}, \textbf{3}, 0) \\ 
&   & &  $ [\equiv \tilde{W}^{\pm} , \tilde{W}^{0}]$ &   &   \\ 
&  &   &      &  &   \\ 
B    & $\mathbf{W_{1\alpha}}$ & & $\lambda_{1\alpha} $                       & $B$              & (\textbf{1}, \textbf{1}, 0 ) \\ 
 &   & & $ [\equiv \tilde{B}]$                       &                &    \\ 
\hline
\hline
DG-octet& $\mathbf{O_g}$ &  $O_g $  & $\chi_{g} $ &  & (\textbf{8}, \textbf{1}, 0) \\ 
&   &  $ [\equiv \Sigma_g]$  &  $  [ \equiv \tilde{g}']$ &  &  \\ 
&  &   &      &  &   \\ 
DG-triplet & $\mathbf{T}$ & $\{T^0, T^{\pm}\}$ & $\{\chi_T^0, \chi_T^{\pm}\}$ &  & (\textbf{1},\textbf{3}, 0 )\\ 
&   &  $[ \equiv \{\Sigma^W_0, \Sigma_W^{\pm}\}]$ & $[ \equiv \{\tilde{W}'^{\pm},\tilde{W}'^{0}\}]$ &  &  \\ 
&  &   &      &  &   \\ 
DG-singlet  & $\mathbf{S}$& $S$ & $\chi_{S} $   &  & (\textbf{1}, \textbf{1}, 0 ) \\ 
&  & $ [\equiv \Sigma_B]$ &  $ [ \equiv \tilde{B}']$    &  &   \\ 
\hline
\end{tabular}
\caption{Chiral and gauge multiplet fields in the model}
\label{diracgauginos_Fields}
\end{center}
\end{table}

The particle content of the model is presented in table \ref{diracgauginos_Fields}. The  MSSM matter fields acquire masses through the Yukawa  superpotential:
\begin{equation}
W_{Yukawa} = y^{ij}_u \mathbf{u^c_i Q_j \cdot H_u} - y^{ij}_d \mathbf{d^c_i Q_j \cdot H_d }
- y^{ij}_e \mathbf{e^c_i L_j \cdot H_d }
\label{diracgauginos_MSSMSuperpotential}
\end{equation}
and the usual soft breaking terms:
\begin{eqnarray}
\mathcal{L}^{0}_{soft} &= &   \tilde{Q}^\dagger_i  {m_Q^2}_{ij} \tilde{Q}_j+ \tilde{L}^\dagger_i  {m_L^2}_{ij} \tilde{L}_j+
\tilde{u}^\dagger_i  {m_u^2}_{ij} \tilde{u}_j+ \tilde{d}^\dagger_i  {m_d^2}_{ij} \tilde{d}_j+ \tilde{e}^\dagger_i  {m_e^2}_{ij} \tilde{e}_j\nonumber \\ && + A^{ij}_u {\tilde{u}^c_i \tilde{Q}_j \cdot H_u} - A^{ij}_d {\tilde{d}^c_i \tilde{Q}_j \cdot H_d }
- A^{ij}_e {\tilde{e}^c_i \tilde{L}_j \cdot H_d } +c.c.
\end{eqnarray}
where the bold characters denote superfields. Here, $i,j$ are family indices and run from $1$ to $3$. The $3 \times 3$ $y$ matrices are the Yukawa couplings. The ``$\cdot$" denotes 
$SU(2)$ invariant couplings, for example: $Q \cdot H_u = \tilde{u}_L H_u^0 - \tilde{d}_L H_u^+ $.

In order to have Dirac masses for the gauginos, additional fields in the adjoint representations, the ``DG-adjoints'', are introduced. 
We define the superfields:
\begin{eqnarray}
\mathbf{S} & = & S + \sqrt{2} \theta \chi_S + \cdots  \\
\mathbf{T} & = & T  + \sqrt{2} \theta \chi_T + \cdots \\
\mathbf{O_g} & = & {O}_g  + \sqrt{2} \theta \chi_g + \cdots
\end{eqnarray}
where $S = \frac{1}{\sqrt{2}}(S_R+i S_I)$ is a singlet and $T= \sum_{a=1,2,3} T^{(a)}$ an $SU(2)$ triplet parametrized as:
\begin{eqnarray}
T^{(1)}= T_1 \frac{\sigma^1}{2},  \qquad  T^{(2)}= T_2 \frac{\sigma^2}{2},&&   T^{(3)}= T_0 \frac{\sigma^3}{2}, \nonumber \\ T=\frac{1}{2} \begin{pmatrix}
T_0     &  \sqrt{2} T_+  \\
\sqrt{2}   T_-  &  -T_0
\end{pmatrix} ,&&  \nonumber \\ 
T_0= \frac{1}{\sqrt{2}}(T_R+i T_I),  \quad T_+=\frac{1}{\sqrt{2}}( T_{+R}+i T_{+I}), && \quad T_-=\frac{1}{\sqrt{2}}( T_{-R}+ iT_{-I}), 
\end{eqnarray}
and $\sigma^a$ are the Pauli matrices. Their quantum numbers are presented in Table \ref{diracgauginos_Fields}.

Due to the presence of these extra fields, the  gauge kinetic terms are modified to become:
\begin{eqnarray}
\mathcal{L}_{gauge}= & &\int d^4x d^2\theta  \left[ \right.   \frac{1}{4} \textbf{M}_{1} \mathbf{W}_{1}^{\alpha}  \mathbf{W}_{1\alpha} + \frac{1}{2}\textbf{M}_{2} \textrm{tr}(\mathbf{W}_{2}^{\alpha}  \mathbf{W}_{2\alpha}) + \frac{1}{2}\textbf{M}_{3} \textrm{tr}(\mathbf{W}_{3}^{\alpha}  \mathbf{W}_{3\alpha}) 
\nonumber \\  && +  \sqrt{2} \textbf{m}^\alpha_{1D} \mathbf{W}_{1\alpha} \mathbf{S}  
+  2\sqrt{2} \textbf{m}^\alpha_{2D} \textrm{tr}(\mathbf{W}_{2 \alpha} \mathbf{T}) +  2\sqrt{2} \textbf{m}^\alpha_{3D} \textrm{tr}(\mathbf{W}_{3 \alpha} \mathbf{O_g}) \left.  \right]\nonumber \\  
&+&\int d^4x d^2\theta d^2{\bar{\theta}}\quad (\sum_{ij}\mathbf{ \Phi}_{i}^\dagger e^{g_j \mathbf{V_j}} \mathbf{\Phi}_{i} + h.c.) 
\label{Newdiracgauge}
\end{eqnarray}
where $\mathbf{V}_j$ are the vector  and $\mathbf{W}_{j \alpha} $ the corresponding field strength superfields associated to $U(1)_Y$, $ SU(2)$ and $ SU(3)$  for $j=1, 2, 3$ respectively. Here, we have introduced  spurion superfields to take into account the generation of gaugino masses:
\begin{eqnarray}
\textbf{M}_{i} &= & 1+ 2 \theta \theta M_{i}\\
\textbf{m}_{\alpha iD} &= & \theta_\alpha m_{iD}
\end{eqnarray}
The Dirac gaugino spurion superfield can be written as  $\textbf{m}^\alpha_{iD} = -\frac{1}{4} \bar {D}\bar{ D}D_\alpha \mathbf{X}_{i} $. This mass originates  as a $D$-term if  $\mathbf{X}_i$  is identified as a vector field $\mathbf{X}_i = \mathbf{V'}/ M_i$, or as non vanishing  $F$-term  by writing $\mathbf{X}_i = 2 \mathbf{\Sigma^\dagger\Sigma}/ M_i$  with $\mathbf{\Sigma} = \theta\theta F$,  where $M_i $ is the appropriate supersymmetry breaking mediation scale.

The DG-adjoints may also modify  the Higgs  superpotential, since new relevant and marginal operators are now allowed:

\begin{equation}
\int d^4x d^2\theta \left[ \mu \mathbf{H_u\cdot H_d }+ \frac {M_S}{2}  \mathbf{S}^2 + \lambda_S \mathbf{SH_d\cdot H_u} + M_T \textrm{tr}(\mathbf{TT}) + 2  \lambda_T \mathbf{H_d\cdot T H_u} \right]
\label{NewSuperPotential}
\end{equation}
with the definition $H_u\cdot H_d = H^+_uH^-_d - H^0_u H^0_d$.

Finally, the  soft supersymmetry breaking terms for the scalars are:
\begin{eqnarray}
- \Delta\mathcal{L}_{soft} &= &   m_{H_u}^2 |H_u|^2 + m_{H_d}^2 |H_d|^2 + B_\mu (H_u\cdot H_d + h.c.) \nonumber \\ &&+ m_S^2  |S|^2 + \frac{1}{2} B_S (S^2 + h.c.)  + 2 m_T^2 \textrm{tr}(T^\dagger T) + B_T (\textrm{tr}(T T)+ h.c.) \nonumber \\ &&+ 
A_S  \lambda_S (SH_d\cdot H_u + h.c.) +  2 A_T  \lambda_T(H_d \cdot T H_u + h.c.)  
\label{Lsoft}
\end{eqnarray}

Note that we did not include in the superpotential a cubic term $\textrm{tr}(\mathbf{TTT})$ as this identically vanishes. Neither did we include linear and cubic terms in the singlet. The latter is due to the fact that we assume that the DG-adjoint appears due to some underlying $N=2$ supersymmetry that forbids these terms. Of course a microscopic model explaining the origin of the soft terms should also address the fact that the supersymmetric $M_S$ and $M_T$ are assumed to take values of order of the electroweak scale,  introducing an issue of scale hierarchy as does the Higgs $\mu$-term. 

Below, we will give special attention to the scenario where the DG-states arise as a result of an $N = 2$ extension of the gauge sector. In this case, if the Higgs multiplets $H_u$ and $H_d$ are assumed to form an $N = 2$ hypermultiplet then  $\lambda_S$ and $ \lambda_T$ are related to the gauge couplings, at the $N=2$ scale, by:
\begin{equation}
\lambda_S= \sqrt{2} g' \frac{1}{2} , \qquad  \lambda_T= \sqrt{2} g \frac{1}{2} , 
\label{Neq2Lambdas}\end{equation}
where $g'$ and $g$ are the  $U(1)_Y$ and $SU(2)$ gauge couplings respectively. The factor $1/2$ in $\lambda_S$ arises from the $U(1)_Y$ charge of the Higgs doublets.

\section{Electroweak scalar potential}
\label{diracgauginos_secInteractions}

We turn now to the electroweak scalar potential. This receives contributions from three sources:
\begin{equation}
{V}_{EW}= {V}_{gauge}+{V}_{W}+{V}_{soft}
\end{equation}
The first is a contribution from  the gauge  kinetic term (\ref{Newdiracgauge}). Integration on the spinor coordinates, and going on-shell, leads to the $U(1)_Y$ and $ SU(2)$ $D$-terms:
\begin{eqnarray}
D_1 & = &  - 2 m_{1D}  S_R +D_Y^{(0)} \qquad \textrm{with}  \qquad D_Y^{(0)}=  - g'\sum_{j} Y_j \varphi_j^* \varphi_j \label{Electroweak:DtermsS}\\
D^a_2 & = &  - \sqrt{2} m_{2D} (T^a+ T^{a\dagger}) +D_2^{a(0)} \qquad \textrm{with}  \qquad D_2^{a(0)}=  -g \sum_{j}  \varphi_j^*  \frac{\sigma^a}{2} \varphi_j 
\label{Electroweak:DtermsT}\end{eqnarray}
where $\varphi_j $ are the scalar components of matter chiral superfields, whereas $D_Y^{(0)}$ and $D_2^{a(0)}$  are the $D$-terms in absence of Dirac masses. The resulting Lagrangian contains terms of the form:

\begin{equation}
\mathcal{L}_{gauge}\rightarrow  -m_{1D} \lambda_1^\alpha   \chi_{S \alpha}    \
- m_{2D} \textrm{tr} ( \lambda_2^\alpha   \chi_{T\alpha}  ) - \frac{1}{2} D_1^2 - \frac{1}{2} \textrm{tr } (D_2^aD_2^a)
\end{equation}
where we can identify the Dirac components of the gauginos  $  \lambda_1 \equiv \tilde{B}'$ and $  \lambda_2^a \equiv \tilde{W}'^a$ as given in Table \ref{diracgauginos_Fields}.

The contribution  from the DG-triplet  to $D_2^{a}$ is:
\begin{equation}
D_2 \propto \frac{1}{2} \begin{pmatrix}
(|T_-|^2 - |T_+|^2)   &  \sqrt{2} ( T_+ T_0^* - T_0T^*_- )  \\
\sqrt{2}   (T_0 T_+^*  - T_-T_0^* )  &  - (|T_-|^2 - |T_+|^2)
\end{pmatrix} 
\end{equation}
which vanishes  in electrically neutral vacuum, where:
\begin{eqnarray}
<T_+ >& = & <T_->=<H_+>=<H_->=0.
\end{eqnarray}
The contribution to the scalar potential of the neutral fields is then given by:
\begin{equation}
V_{gauge}  =   2 m^2_{1D}  S^2_R -     2 m_{1D}  S_R    D_Y^{(0)}  +  \frac{1}{2}   D_Y^{(0)2}
+  2 m^2_{2D}  T^2_R -     2 m_{2D}  T_R    D_2^{(0)}  +  \frac{1}{2}  D_2^{(0)2} 
\end{equation}
where we have dropped the generator  label, $D_2= D_2^{a}, a=3 $, for the only non-vanishing component.

The second contribution comes from the superpotential (\ref{NewSuperPotential}):
\begin{eqnarray}
W & = &  (-\mu + \lambda_S \mathbf{S}) \mathbf{(H^0_u H^0_d- H^+_uH^-_d )} +  \frac {M_S}{2} \mathbf{S}^2 + \frac{M_T}{2}  (\mathbf{T^0 T^0 + 2 T^+ T^-}   )\nonumber \\
& &   - \lambda_T  \mathbf{( H^0_d T^0 H^0_u + H^-_d T^0 H^+_u)} - \sqrt{2}  \lambda_T  \mathbf{ ( H^-_d T^+ H^0_u - H^0_d T^- H^+_u) }.
\end{eqnarray}
Keeping only the neutral components, it reads:
\begin{equation}
V_W  =   |M_S S + \lambda_S H^0_d H^0_u|^2 + |M_T T^0 - \lambda_T H^0_d H^0_u|^2 +  |\mu - \lambda_S S + \lambda_T T^0|^2 
(|H^0_d|^2+|H^0_u|^2)
\end{equation}

The third source ${V}_{soft}$ is due to soft supersymmetry breaking terms for the scalars (\ref{Lsoft}).
We define:
\begin{equation}
H^0_u  = \frac{H^0_{uR} + i H^0_{uI}}{\sqrt{2}}  , \quad H^0_d  = \frac{H^0_{dR} + i H^0_{dI}}{\sqrt{2}}
\end{equation}
then, all together, the scalar Lagrangian for the neutral fields is now given by:
\begin{eqnarray}
V_{EW} &= & 
(m_{H_u}^2 +\mu^2) |H^0_u|^2 + (m_{H_d}^2 +\mu^2)|H^0_d|^2 - B_\mu (H_u^0H_d^0 + h.c.) + \frac{g^2+g'^2}{8} ( |H_u^0|^2-|H_d^0|^2)^2              \nonumber \\ 
&&+( \lambda_S^2 +  \lambda_T^2) |H_u^0H_d^0|^2  \\ 
&&+
\frac{1}{2} (M_S^2+m_S^2+4 m^2_{1D}+ B_S)  S_R^2 +  \frac{1}{2} (M_S^2+m_S^2-B_S) S_I^2  \nonumber \\ 
&&+
\frac{1}{2} (M_T^2+m_T^2+4 m^2_{2D}+B_T)  T_R^2 +  \frac{1}{2} (M_T^2+m_T^2-B_T) T_I^2 
\nonumber \\ 
&&
+\left[ \frac{\ \lambda_S^2}{2} (S_R^2  + S_I^2 )  + \frac{\ \lambda_T^2}{2} (T_I^2  + T_R^2 ) -{\sqrt{2}} \mu (   \lambda_S S_R  - \lambda_T T_R  )-
\lambda_S \lambda_T ( S_I T_I+  S_R T_R ) \right] \nonumber \\
&& \qquad\times \left[ |H^0_u|^2 + |H^0_d|^2\right]
\nonumber \\ &&
+  g'    m_{1D}  S_R   ( |H_u^0|^2-|H_d^0|^2)   
+   g  m_{2D}  T_R     ( |H_d^0|^2-|H_u^0|^2) 
\nonumber \\ &&
+\frac{\lambda_S }{\sqrt{2}}(  M_S+    A_S)S_R (H^0_{dR} H^0_{uR} -   H^0_{dI} H^0_{uI} )
+ \frac{\lambda_S }{\sqrt{2}}( M_S-   A_S)S_I ( H^0_{dR} H^0_{uI} + H^0_{dI} H^0_{uR})
\nonumber \\ &&
- \frac{\lambda_T }{\sqrt{2}} (  M_T+    A_T)T_R( H^0_{dR} H^0_{uR} -   H^0_{dI} H^0_{uI} )
- \frac{\lambda_T }{\sqrt{2}} ( M_T-   A_T)T_I( H^0_{dR} H^0_{uI} +H^0_{dI} H^0_{uR})\nonumber
\label{VEW}\end{eqnarray}

Note that the MSSM potential is given by the first line. All the parameters are chosen to be real.
We are left with four neutral fields $S_R, S_I,  T_R, T_I$ satisfying equations of type:
\begin{equation}
M^2_{xa} x_a + X_{ST} y_a  =  V_{xa}, \qquad  a= R,I \qquad  \textrm{for}  \   \ (x=S, y=T)  \   \  \textrm{or}  \   \  (x=T, y=R)
\end{equation}
with solutions of the form:
\begin{equation}
x_a= \frac{V_{xa} M^2_{ya}-V_{ya} X_{ST} }{M^2_{xa}M^2_{ya}-X_{ST}^2},\qquad  a= R,I \qquad  \textrm{for}  \   \ (x=S, y=T)  \   \  \textrm{or}  \   \  (x=T, y=R)
\end{equation}

where:
\begin{eqnarray}
M^2_{SR}& = & M_S^2+m_S^2+4 m^2_{1D}+ B_S + \lambda_S^2 ( |H^0_u|^2 + |H^0_d|^2) \\
M^2_{TR}& = & M_T^2+m_T^2+4 m^2_{2D}+ B_T + \lambda_T^2 ( |H^0_u|^2 + |H^0_d|^2) \\
M^2_{SI}& = & M_S^2+m_S^2- B_S + \lambda_S^2  ( |H^0_u|^2 + |H^0_d|^2) \\
M^2_{TI}& = & M_T^2+m_T^2-B_T + \lambda_T^2 ( |H^0_u|^2 + |H^0_d|^2) 
\end{eqnarray}
while
\begin{equation}
X_{ST} = - \lambda_S  \lambda_T  ( |H^0_u|^2 + |H^0_d|^2) 
\end{equation}
and:
\begin{eqnarray}
V_{SR}& = & {\sqrt{2}} \mu    \lambda_S ( |H^0_u|^2 + |H^0_d|^2) -g'    m_{1D}  ( |H_u^0|^2-|H_d^0|^2)   \nonumber \\ 
&& - \frac{\lambda_S }{\sqrt{2}}(  M_S+    A_S) (H^0_{dR} H^0_{uR} -   H^0_{dI} H^0_{uI} )  \\
V_{TR}& = & -{\sqrt{2}} \mu    \lambda_T ( |H^0_u|^2 + |H^0_d|^2) +g   m_{2D}  ( |H_u^0|^2-|H_d^0|^2)  \nonumber \\ 
&& +  \frac{\lambda_T }{\sqrt{2}} (  M_T+    A_T)( H^0_{dR} H^0_{uR} -   H^0_{dI} H^0_{uI} )\\
V_{SI}& = & -\frac{\lambda_S }{\sqrt{2}}( M_S-   A_S)( H^0_{dR} H^0_{uI} + H^0_{dI} H^0_{uR})   \\
V_{TI}& = &+  \frac{\lambda_T }{\sqrt{2}} ( M_T-   A_T)( H^0_{dR} H^0_{uI} +H^0_{dI} H^0_{uR})
\end{eqnarray}

We are not going to pursue exact computations. Instead, we will consider the case with $m^2_S, m_T^2 \gg m_Z^2$  in which case the formulae simplify as we can neglect all terms proportional to $ ( |H^0_u|^2 + |H^0_d|^2)$, in particular  $X_{ST}$. This gives $x_a \simeq {V_{xa}  }/{M^2_{xa}}$, i.e.
\begin{eqnarray}
S_R \!&\simeq &\!- \frac{ g'    m_{1D}  ( |H_u^0|^2-|H_d^0|^2) - {\sqrt{2}} \mu    \lambda_S ( |H^0_u|^2 + |H^0_d|^2)  + \frac{\lambda_S }{\sqrt{2}}(  M_S+    A_S) (H^0_{dR} H^0_{uR} -   H^0_{dI} H^0_{uI} ) }{ M_S^2+m_S^2+4 m^2_{1D}+ B_S} \nonumber \\
T_R \!&\simeq &\!- \frac{ g    m_{2D}  ( |H_d^0|^2-|H_u^0|^2) +{\sqrt{2}} \mu    \lambda_T ( |H^0_u|^2 + |H^0_d|^2)- \frac{\lambda_T }{\sqrt{2}} (  M_T+    A_T)( H^0_{dR} H^0_{uR} -   H^0_{dI} H^0_{uI} ) }{ M_T^2+m_T^2+4 m^2_{2D}+ B_T}\nonumber \\
S_I \!&\simeq  &\!-\lambda_S \frac{( M_S-   A_S) ( H^0_{dR} H^0_{uI} +H^0_{dI} H^0_{uR}) }{{\sqrt{2}}( M_S^2+m_S^2+4 m^2_{1D}+ B_S)}\nonumber \\
T_I\! &\simeq &\! \lambda_T \frac{( M_T-   A_T) ( H^0_{dR} H^0_{uI} +H^0_{dI} H^0_{uR})  }{{\sqrt{2}}(M_T^2+m_T^2+4 m^2_{2D}+ B_T)}
\end{eqnarray}

We are interested by the case of CP neutral vacuum, i.e. $H^0_{dI}=H^0_{uI} =0$ which implies $S_I=T_I=0$. From now on, we will drop the indices $0$ and define\footnote{With this convention $v \simeq 246$ GeV, $\frac{(g')^2 + g^2}{4} v^2 = M_Z^2$.}:
\begin{eqnarray}
<H^0_{uR}>\equiv <h_u>&=& {v_u}= {v s_\beta }, \quad <H^0_{dR}>\equiv <h_d>  = {v_d}={v c_\beta}, \qquad   0 \leqslant \beta \leqslant \frac{\pi}{2} \nonumber \\
<S_R>&=& v_s \qquad <T_R>=v_t
\end{eqnarray}
where we denote:
\begin{eqnarray}
c_\beta &\equiv& \cos \beta,\qquad   s_\beta \equiv \sin \beta, \qquad  t_\beta \equiv \tan\beta \nonumber \\
c_{2\beta} &\equiv& \cos 2\beta,\qquad   s_{2\beta} \equiv \sin 2\beta
\end{eqnarray}

\begin{eqnarray}
v_s &\simeq &\frac{v^2}{2(M_S^2+m_S^2+4 m^2_{1D}+ B_S)} \ \ { \left[g'    m_{1D} c_{2\beta} + {\sqrt{2}} \mu    \lambda_S   - \frac{\lambda_S }{\sqrt{2}}(  M_S+    A_S)  s_{2\beta}  \right]} \nonumber \\
v_t & \simeq &  \frac{v^2}{2(M_T^2+m_T^2+4 m^2_{2D}+ B_T)} \ \ \left[ - g    m_{2D}  c_{2\beta} -{\sqrt{2}} \mu    \lambda_T + \frac{\lambda_T }{\sqrt{2}} (  M_T+    A_T)  s_{2\beta}  \right].
\end{eqnarray}
Electroweak precision data give strong bounds on the expectation value of the DG-triplet as it contributes to $\rho \simeq 1 +\alpha T = 1.0004^{+0.0008}_{-0.0004}$~\cite{Amsler:2008zzb}. Thus we require:
\begin{equation}
\Delta \rho  \simeq 4 \frac{v_t^2}{v^2} \lesssim 8 \cdot 10^{-4}
\end{equation}
which is satisfied for $v_t\lesssim 3$ GeV. Here we will allow the Dirac and Majorana masses to vary arbitrarily and will satisfy this bound by taking $m_T$ large enough. For instance, for $M_S, A_S, \mu \sim 200$ GeV and all couplings of order one,  this is satisfied for $m_T\gtrsim 1$ TeV.

Integrating out the DG-adjoints $S$ and $T$, leads then to the effective tree-level scalar potential at first order in $\lambda_{S,T}$
\begin{eqnarray}
V_{EW} &= & 
\frac{(m_{H_u}^2 +\mu^2)}{2} h_u^2 + \frac{(m_{H_d}^2 +\mu^2)}{2} h_d^2 -  {B_\mu} h_uh_d + \frac{g^2+g'^2}{32} ( h_u^2-h_d^2)^2              \nonumber \\ 
&&+ \frac{ \lambda_S^2 +  \lambda_T^2}{4} h_u^2 h_d^2 \nonumber \\ 
&&-
\frac{1}{8} 
\frac{ \left[g'    m_{1D}  ( h_u^2-h_d^2) - {\sqrt{2}} \mu    \lambda_S ( h_u^2 + h_d^2)  + {\sqrt{2}}{\lambda_S }(  M_S+    A_S) h_{d} h_{u}  \right]^2}{ M_S^2+m_S^2+4 m^2_{1D}+ B_S}   \nonumber \\ 
&&-
\frac{1}{8} \frac{  \left[ -g    m_{2D}  ( h_u^2-h_d^2) +{\sqrt{2}} \mu    \lambda_T ( h_u^2 + h_d^2)- {\sqrt{2}}  {\lambda_T }(  M_T+    A_T) h_{d} h_{u} \right]^2 }{ M_T^2+m_T^2+4 m^2_{2D}+ B_T}  \nonumber \\ 
\end{eqnarray}

This decomposes into three parts:
\begin{equation}
V_{EW}= V_0 +V_1 + V_2
\end{equation}
The first part:
\begin{equation}
V_0 = \frac{(m_{H_u}^2 +\mu^2)}{2} h_u^2 + \frac{(m_{H_d}^2 +\mu^2)}{2} h_d^2 - {B_\mu} h_uh_d + \frac{g^2+g'^2}{32} ( h_u^2-h_d^2)^2  
\end{equation}
is the MSSM contribution. The second,
\begin{equation}
\label{newquartic}
V_1 = \frac{ \lambda_S^2 +  \lambda_T^2}{4} h_u^2h_d^2
\end{equation}
is a quartic term. The third contains  the explicit dependence on the mass parameters of the DG-adjoints. We will illustrate this in taking a few limits, keeping the other parameters fixed:
\begin{itemize}
\item One limit is to take $M_S \rightarrow \infty$ and $M_T \rightarrow \infty$. In this case:
\begin{equation}
V_2\longrightarrow - \frac{ \lambda_S^2 +  \lambda_T^2}{4} h_u^2h_d^2 = -V_1
\end{equation}
meaning that the complete DG-adjoint supermultiplets have been decoupled and one is left with the MSSM electroweak scalar potential.

\item A second limit is to take $\lambda_S \rightarrow 0$ and $\lambda_T \rightarrow 0$ switching off the superpotential couplings
\begin{eqnarray}
V_1 &\longrightarrow& 0 \\
V_2 &\longrightarrow & -\left[\frac{g'^2}{32} (\frac{4 m^2_{1D}}{4 m^2_{1D} +m_S^2 + M_S^2+B_S} )+\frac{g^2}{32}( \frac{4 m^2_{2D}}{4 m^2_{2D} +m_T^2 + M_T^2+B_T})\right] ( h_u^2-h_d^2)^2 \nonumber 
\end{eqnarray}
This shows that the effect of Dirac masses is to decrease the quartic coupling, and  make it vanish in the absence of other  masses.

\item A third one is to take $m_{1D} \rightarrow \infty$ and $m_{2D} \rightarrow \infty$
\begin{eqnarray}
V_1 &\longrightarrow& \frac{ \lambda_S^2 +  \lambda_T^2}{4} h_u^2h_d^2 \nonumber \\
V_2 &\longrightarrow & -\frac{g^2+g'^2}{32} ( h_u^2-h_d^2)^2 
\end{eqnarray}
the Higgs quartic term of the MSSM due to $D$-term is cancelled and all the quartic couplings are generated by superpotential coupling between the Higgses and the DG-adjoints.

\item As a fourth one, we consider the case of interest in the rest of this work: $m_S \rightarrow \infty$ and $m_T \rightarrow \infty$
\begin{eqnarray}
V_1 &\longrightarrow&  \frac{ \lambda_S^2 +  \lambda_T^2}{4} h_u^2h_d^2 \nonumber \\
V_2 &\longrightarrow & 0
\end{eqnarray}
showing that the MSSM scalar potential is supplemented with a quartic term that lifts the $D$-term flat direction $H_u = H_d$.
\end{itemize}

\subsection{Integrating out the adjoints}

In this last limit, the vacuum expectation values of the DG-adjoints can be neglected, allowing to write an $SU(2)$ invariant effective scalar potential. It can be put in the usual form \cite{Haber:1993an,Boudjema:2001ii} which parametrizes the two-doublet potential:

\begin{eqnarray}
\label{potential4}
V_{eff} & = & (m_{H_u}^2+\mu^2) |H_u|^2 +
(m_{H_d}^2+\mu^2) |H_d|^2
            - [m_{12}^2 H_u\cdot H_d + h.c. ] \nonumber \\
      &   & + \frac{1}{2}\big[\frac{1}{4}(g^2+g'^2) + \lambda_1\big]
                          (|H_d|^2)^2
            +\frac{1}{2}\big[\frac{1}{4}(g^2+g'^2) + \lambda_2\big]
                          (|H_u|^2)^2 \nonumber \\
      &  &  +\big[\frac{1}{4}(g^2-g'^2) + \lambda_3\big]
                          |H_d|^2 |H_u|^2
            +\big[-\frac{1}{2}g^2 + \lambda_4\big]
(H_d\cdot H_u)(H_d^*\cdot H_u^*)\nonumber\\
      & &   +\big(\frac{\lambda_5}{2} ( H_d\cdot H_u)^2
                 +\big[ \lambda_6 |H_d|^2 + \lambda_7 |H_u|^2\big]
                       ( H_d\cdot H_u) + h.c. \big)
\end{eqnarray}
where now:
\begin{eqnarray}
\lambda_3 &=& 2 \lambda^2_T \qquad \lambda_4= \lambda^2_S-\lambda^2_T \nonumber\\
\lambda_1&=&\lambda_2=\lambda_5= \lambda_6=\lambda_7=0.
\end{eqnarray}

As in the MSSM we expect sizeable one-loop corrections to the above potential, but we will also have new and particularly important corrections due to taking $m_S, m_T$ large. We can approximate these new contributions by taking the leading logarithmic contributions from scalar loops involving only quartic vertices; the remaining diagrams involving scalars and sfermions will have the effect of removing the dependence on the cutoff. We can reasonably approximate this behaviour by replacing the renormalisation scale in the logarithms with $v$ (a more precise calculation involves the full fermion mass matrices). We should then also evaluate the coupling constants at this scale. We can then write

\begin{eqnarray}
\delta^{(1)} \lambda_1 &=& \frac{3}{16\pi^2} y_b^4 \log \left(\frac{m_{\tilde{b}_1}m_{\tilde{b}_2} }{v^2} \right) + \frac{5}{16\pi^2} \lambda_T^4 \log \left(\frac{m_T^2}{v^2}\right) + \frac{1}{16\pi^2} \lambda_S^4 \log \left(\frac{m_S^2}{v^2}\right) \nonumber \\
&&- \frac{1}{16\pi^2} \frac{\lambda_S^2 \lambda_T^2}{m_T^2 - m_S^2} \bigg\{ m_T^2 [\log \left(\frac{m_T^2}{v^2}\right) -1] - m_S^2 [\log \left(\frac{m_S^2}{v^2}\right) -1] \bigg\}\nonumber \\
\delta^{(1)} \lambda_2 &=& \frac{3}{16\pi^2} y_t^4 \log \left(\frac{m_{\tilde{t}_1}m_{\tilde{t}_2} }{m_t^2} \right) +\frac{5}{16\pi^2} \lambda_T^4 \log \left(\frac{m_T^2}{v^2}\right) + \frac{1}{16\pi^2} \lambda_S^4 \log \left(\frac{m_S^2}{v^2}\right) \nonumber \\
&&- \frac{1}{16\pi^2} \frac{\lambda_S^2 \lambda_T^2}{m_T^2 - m_S^2} \bigg\{ m_T^2 [\log \left(\frac{m_T^2}{v^2}\right) -1] - m_S^2 [\log \left(\frac{m_S^2}{v^2}\right) -1] \bigg\}\nonumber \\ 
\delta^{(1)} \lambda_3 &=& \frac{5}{32\pi^2} \lambda_T^4 \log \left(\frac{m_T^2}{v^2}\right) + \frac{1}{32\pi^2} \lambda_S^4 \log \left(\frac{m_S^2}{v^2}\right) \nonumber \\
&&+ \frac{1}{32\pi^2} \frac{\lambda_S^2 \lambda_T^2}{m_T^2 - m_S^2} \bigg\{ m_T^2 [\log \left(\frac{m_T^2}{v^2}\right) -1] - m_S^2 [\log \left(\frac{m_S^2}{v^2}\right) -1] \bigg\}
\end{eqnarray} 
and can neglect the contributions to the other factors as subleading. For simplicity we have assumed $B_S, B_T \ll m_S^2, m_T^2$; this is not valid in the scenario of for example \cite{Benakli:2008pg} where they are of equal order in magnitude, both being generated at one loop. 

Note that there is also a contribution to the Higgs mass parameters proportional to \\*$\lambda_{S,T}^2 m_{S,T}^2 \log \left(\frac{m_{S,T}^2}{v^2}\right)$, which is absorbed into the renormalisation of $m_{H_u}^2, m_{H_d}^2$. If these are large then we have a large fine tuning.

\subsection{Singlet Extension to the Higgs Sector}

As can be seen from equations (\ref{Electroweak:DtermsS},\ref{Electroweak:DtermsT}), the sfermions obtain a mass proportional to the expectation values of $S_R, T_R$. Of these, we can ignore the $T_R$ contribution due to the strong constraints upon it as discussed above, but $\langle S_R\rangle$ may be non-negligible. Together with the dominant one loop effect (containing gauginos or $S_R, T_R$ scalars) we find
\begin{equation}
m_{ii}^2 = (m_{ii}^{(0)})^2 + 2 m_{1D} \langle S_R \rangle g' Y_i + Y_i^2 (g')^2 \frac{m_{1D}^2}{4\pi^2} \log \frac{m_{S_R}^2}{m_{\lambda_1}^2} + \frac{1}{2} g^2 \frac{m_{2D}^2}{4\pi^2} \log \frac{m_{T_R}^2}{m_{\lambda_2}^2}
\end{equation}
where, since the action that we are using is an effective one, we have included the soft masses induced by the supersymmetry breaking sector (via gauge mediation or otherwise) other than through the Dirac gauginos as $m_{ii}^{(0)}$. For example, in gauge mediation these are generated at two loops; the gaugino masses $m_{1D}, m_{2D}, m_{\lambda_1},m_{\lambda_2}$ (the latter two are the total gaugino mass, including both Dirac and any Majorana effects, defined as the location of the pole in the $\langle \lambda \overline{\lambda}\rangle$ propagator) are generated at one. The one loop fluctuations around our effective action then actually appear in gauge mediation at three loops. 

The term proportional to $\langle S_R \rangle$ clearly gives a negative contribution to negatively charged states, so we must ensure that this does not dominate. This is particularly important for models where $m_{ii}^{(0)}$ vanishes, which can occur for certain gauge mediation models with purely Dirac gauginos (in such cases $M_S, M_T = 0$). One way to ensure positivity is to take a large $m_S, m_T$, which reduces $\langle S_R \rangle$ and increases the loop effects. We then have (where $m_{S_R, T_R}^2 = m_{S,T}^2 + B_{S,T}$)
\begin{equation}
\!\frac{m_{1D} g' Y_i v^2}{2} \bigg(\frac{  c_{2\beta} g'  m_{1D}  + \sqrt{2}\mu    \lambda_S  - 2 \sqrt{2}\lambda_S  A_S s_\beta c_\beta  }{ m_S^2+4 m^2_{1D}+ B_S}\bigg)\! + \!Y_i^2 (g')^2 \frac{m_{1D}^2}{4\pi^2} \log \frac{m_{S_R}^2}{m_{\lambda_1}^2}\!+\!\frac{g^2}{2}  \frac{m_{2D}^2}{4\pi^2} \log \frac{m_{T_R}^2}{m_{\lambda_2}^2}\!>\!0.\! 
\nonumber\end{equation}
This is easy to arrange for large $m_{S_R}$, for example for small $\lambda_S$ or large $m_{1D}c_{2\beta}  \ll m_S$ we require (from $SU(2)$ singlets, which provide the strictest constraint)
\begin{equation}
\log (m_S^2 +B_S) / m_{\lambda_1}^2 > \frac{2\pi^2 v^2}{m_S^2 + B_S + 4 m_{1D}^2}
\end{equation}
and therefore for $m_{1D}^2 \sim v^2$ we have $m_{S_R} \gtrsim 3v + 0.94(m_{\lambda_1}-v) + ... \gtrsim 750 GeV$, and for $m_{1D}^2 =   x m_{S_R}^2 $ we require $m_{S_R} \gtrsim \pi v/\sqrt{-(2x+1/2)\log x}$. If we consider for example that the masses are generated in gauge mediation, then $m_{1D} \sim \lambda_X g' \Lambda, m_S \sim B_S \sim \lambda_X^2 \Lambda^2$ for a messenger coupling $\lambda_X$ and effective supersymmetry breaking scale $\Lambda$; this gives $x = (g')^{2} \sim 1/8.4$ and thus $m_{S_R} \gtrsim 620 GeV$. However, this limit is only valid for $\lambda_S \ll  g'/\sqrt{2}$, as the gaugino masses are smaller than $v$. For $\lambda_S \sim g'/\sqrt{2}$ we can place a bound by setting $\mu \sim v$:
\begin{equation}
-\frac{\sqrt{x} (g')^2 v^3}{2m_{S_R}(4x+1)} -  (g')^2 \frac{x m_{S_R}^2}{4\pi^2} \log x > 0
\end{equation}   
and thus $m_{S_R} > v\bigg(-2\pi^2/(\sqrt{x}(1+4x)\log x)\bigg)^{1/3} \gtrsim 650 GeV$.

Now consider when $m_S$ is small, $\lambda_S \langle S_R \rangle \equiv \tilde{\mu}$ is large and adds to or replaces the $\mu$ term while $A_S \lambda_S \langle S_R \rangle$ contributes to $B\mu$. The loop terms are no longer significant, and we must simply impose that
\begin{equation}
(m_{ii}^{(0)})^2  + 2 m_{1D} \frac{\tilde{\mu}}{\lambda_S} g' Y_i > 0.
\end{equation}
This will be in general quite a stringent constraint on the parameter space of the microscopic model.

Keeping $S$ in the light spectrum results in the effective potential (\ref{VEW}) with $T=0$. The MSSM spectrum is extended by one CP even Higgs and one CP odd state, corresponding to $S_R$ and $S_I$ respectively. This falls in the class of models studied by~\cite{Chang:2005ht}. In addition to the possibility discussed there of new decays of the lightest Higgs into pairs of CP odd states (as can be seen from Eq.~\ref{VEW}) we would like to stress the new feature that the Higgs has different decay/production channels due to the mixing with $S_R$ which couples to the D term as 
\begin{equation}
\mathcal{L}_{int} \supset - 2 m_{1D}  S_R g'\sum_{j} Y_j \varphi_j^* \varphi_j .
\end{equation}
The effect of this at colliders is very model dependent. It is important, for light sfermions, when $m_{S_R}$ is smaller than or comparable to 
$m_{1D}$. For example, if sneutrinos are arranged to be light then the Higgs may decay to them and then to neutrinos, plus the LSP.

\section{Fermion mass matrix}
\label{FermionicMasses}

\subsection{The neutralinos}
\label{diracgauginos_secNeutralino}

There are six  neutral fermions of interest: the
higgsinos $\tilde{H}^0_u$ and $\tilde{H}^0_d$, the gauginos, bino $\tilde{B}$ and wino $\tilde{W}^0$ and the DG-adjoint fermions 
$\tilde{B}'$ and wino $\tilde{W}'^0$. The mass terms for these fields  have different origins:
\begin{itemize}
\item  bino, wino and DG-adjoints Majorana masses:
\begin{equation}
-\frac{1}{2} ( M_2 \tilde{W}^0 \tilde{W}^0 
+  M_1 \tilde{B} \tilde{B} + M'_2 \tilde{W}'^{0} \tilde{W}'^{0} 
+ M'_1 \tilde{B}' \tilde{B}'  + h.c.) .
\label{diracgauginos_NeutralinoMasses1}
\end{equation}

\item  bino and winos Dirac masses:
\begin{equation}
- m_{2D} \tilde{W}^{\alpha} \tilde{W}'^{\alpha} 
-  m_{1D} \tilde{B} \tilde{B}' 
+ h.c.
\label{diracgauginos_ExtendedSoftMasses}
\end{equation}

\item  The gauge interactions between   the gauginos, the higgsinos and the scalar Higgs:
\begin{equation}
-  \frac{g'}{\sqrt{2}} \left( H^*_u \sigma^i \tilde{H}_u \tilde{B} 
- H^*_d \sigma^i \tilde{H}_d \tilde{B}  \right) 
- \frac{g}{\sqrt{2}} \left( H^*_u \sigma^i \tilde{H}_u \tilde{W}^i 
+ H^*_d \sigma^i \tilde{H}_d \tilde{W}^i  \right) 
\label{diracgauginos_MSSMInteraction}
\end{equation}
leading to 
\begin{equation}
- m_Z \left[ 
\sin \theta_W ( s_\beta \,  \tilde{H}^0_u \tilde{B} 
-  c_\beta  \, \tilde{H}^0_d \tilde{B} )
+\cos \theta_W  ( c_\beta \,  \tilde{H}^0_d \tilde{W}^0
- s_\beta \, \tilde{H}^0_u \tilde{W}^0 ) 
+ h.c.\right]
\label{diracgauginos_NeutralinoMasses2}
\end{equation}

\item The superpotential  (\ref{NewSuperPotential}) leads to couplings between the DG-adjoint fermions, Higgs  and  Higgsinos  
\begin{equation}
- \lambda_S \left( H_d \cdot \tilde{H}_u  \tilde{B}' 
- H_u \cdot \tilde{H}_d \tilde{B}'   \right)
- \lambda_T  \left[ H_u \cdot ( \sigma^i \tilde{H}_d ) \tilde{W}'^i 
+ H_d \cdot ( \sigma^i \tilde{H}_u ) \tilde{W}'^i   \right] 
\label{diracgauginos_ExtendedInteraction}
\end{equation}

giving
\begin{equation}
- m_Z \left[ 
\frac{ \sqrt{2} \lambda_S \sin \theta_W}{g'} ( s_\beta \tilde{H}^0_d \tilde{B}' 
+  c_\beta \tilde{H}^0_u \tilde{B}')
- \frac{ \sqrt{2} \lambda_T \cos \theta_W }{g}  ( c_\beta \tilde{H}^0_u \tilde{W}'^0
+ s_\beta \tilde{H}^0_d \tilde{W}'^0  )
+ h.c.\right]
\label{diracgauginos_ExtendedNeutralinoMasses}
\end{equation}

\item The $\mu$ term in the superpotential $W$  contributes 
to the higgsinos masses ,
\begin{equation}
\mu \tilde{H}^0_u \cdot \tilde{H}^0_d + h.c.
\label{diracgauginos_NeutralinoMasses3}
\end{equation}
\end{itemize}

All the previous terms together describe the resulting mass matrices for 
both neutral  gauginos  and higgsinos when both Majorana and Dirac term are present.  The 
neutralino mass matrix $\mathcal{M}_0$, in the 
$(\tilde{B}', \tilde{B}, \tilde{W}'^0, \tilde{W}^0, \tilde{H}^0_d, \tilde{H}^0_u)$ basis is:
\begin{equation}
\left(\begin{array}{c c c c c c}
M'_1  & m_{1D} & 0     & 0     &  \frac{ \sqrt{2} \lambda_S }{g'}m_Z s_W s_\beta &    \frac{ \sqrt{2} \lambda_S }{g'}m_Z s_W c_\beta  \\
m_{1D} & M_1   & 0     & 0     & -m_Z s_W c_\beta &   m_Z s_W s_\beta  \\
0     & 0     & M'_2  & m_{2D} & - \frac{ \sqrt{2} \lambda_T  }{g}m_Z c_W s_\beta & - \frac{ \sqrt{2} \lambda_T  }{g}m_Z c_W c_\beta  \\
0     & 0     & m_{2D} & M_2   &  m_Z c_W c_\beta & - m_Z c_W s_\beta  \\
\frac{ \sqrt{2} \lambda_S }{g'}m_Z s_W s_\beta & -m_Z s_W c_\beta & -\frac{ \sqrt{2} \lambda_T  }{g}m_Z c_W s_\beta &  m_Z c_W c_\beta & 0    & -\mu \\
\frac{ \sqrt{2} \lambda_S }{g'}m_Z s_W c_\beta &  m_Z s_W s_\beta & -\frac{ \sqrt{2} \lambda_T  }{g}m_Z c_W c_\beta & -m_Z c_W s_\beta & -\mu & 0    \\
\end{array}\right) 
\label{diracgauginos_NeutralinoMassarray}
\end{equation}
where $c_W = \cos \theta_W$, $s_W = \sin \theta_W$. This may be diagonalised by a unitary matrix $N$ such that $\mathcal{M}^{diag}_0=N^* \mathcal{M}_0 N^\dagger$; perturbative expansions for $N$ in various limits are given in appendices \ref{Appendix:BinoWino} and  \ref{Appendix:Higgsino}.

\subsection{The charginos}
\label{diracgauginos_secExtendedCharginos}

The chargino mass matrix describes the mixing between the charged higgsinos $\tilde{H}^+_u$, $\tilde{H}^-_d$ and the charged 
gauginos $\tilde{W}^+$ and $\tilde{W}^-$. The corresponding mass terms have the same origin as 
those presented in section \ref{diracgauginos_secNeutralino}.  There are Dirac masses:
\begin{equation}
- M_2 \tilde{W}^+ \tilde{W}^-  -m_{2D} \tilde{W}^+ \tilde{W}'^- + h.c.
\label{diracgauginos_CharginoMasses1}
\end{equation}
There is the usual $\mu$-term
\begin{equation}
- \mu \tilde{H}^+_u \cdot \tilde{H}^-_d + h.c.
\label{diracgauginos_CharginoMasses3}
\end{equation}
and finally there are mixing terms of the gauginos with the Higgsinos:
\begin{equation}
- \sqrt{2} m_W \sin \beta \tilde{H}^+_u \tilde{W}^- 
- \sqrt{2} m_W \cos \beta \tilde{H}^-_d \tilde{W}^+ + h.c.
\label{diracgauginos_CharginoMasses2}
\end{equation}
and with the DG-adjoints
\begin{equation}
- \frac{ {2} \lambda_T  }{g} m_W c_\beta \tilde{H}^+_u \tilde{W}'^- 
+ \frac{ {2} \lambda_T  }{g}  m_W s_\beta \tilde{H}^-_d \tilde{W}'^+ + h.c.
\label{diracgauginos_ExtendedCharginoMasses}
\end{equation}

The mass terms for the charginos can be expressed in the form 
\begin{equation}
- \frac{1}{2} ( (v^-)^T M_{Ch} v^+ + (v^+)^T M_{Ch}^T v^- + h.c)
\label{diracgauginos_CharginoMassLagrangian}
\end{equation}
where we have adopted the basis $v^+ = (\tilde{W}'^+,\tilde{W}^+,\tilde{H}^+_u)$, 
$v^- = (\tilde{W}'^-,\tilde{W}^-,\tilde{H}^-_d)$. 
Collecting all the terms presented in equations (\ref{diracgauginos_CharginoMasses1})-(\ref{diracgauginos_CharginoMasses3}), 
(\ref{diracgauginos_ExtendedSoftMasses}) and  (\ref{diracgauginos_ExtendedCharginoMasses})  leads to the 
chargino mass matrix :
\begin{equation}
M_{Ch} = 
\left(\begin{array}{c c c}
M'_2   & m_{2D} &\frac{ {2} \lambda_T  }{g} m_W c_\beta \\
m_{2D} & M_2   & \sqrt{2} m_W s_\beta \\
- \frac{ {2} \lambda_T  }{g} m_W s_\beta & \sqrt{2} m_W c_\beta & \mu \\
\end{array}\right) .
\label{diracgauginos_CharginoMassarray}
\end{equation}
This nonsymmetric matrix is diagonalized by separate unitary transformations in 
the basis $v^+$ and $v^-$, $M_{Ch}^{diag} = U^{\dagger} M_{Ch} V$, where  $U$ and $V$ are unitary.

\section{The LSP dark matter}

The model has $R$-parity so the LSP is stable. Here, we assume the LSP to be the (lightest) neutralino and we would like to study its relic density and see how it fits with actual bounds from WMAP. However, before doing so we would like to make some comments on the fate of gravitinos in this scenario. Gravitinos play two major roles in our model: 1) there is the issue of $R$-symmetry breaking by the gravitino mass $m_{3/2}$ and inducing  Majorana masses for the gauginos 2) the gravitino may be the LSP for instance in models of gauge mediation. Depending on its mass it can also change the LSP relic abundance.

First, let us discuss the first issue. In $N=1$ supergravity the gravitino mass needs to be non-zero to cancel the cosmological constant after supersymmetry breaking. It is proportional to the vev of the superpotential and thus breaks $R$-symmetry. If one insists on avoiding the breaking of $R$ symmetry one option is to enhance the gravitational sector to $N=2 (\equiv N=1_1\oplus N=1_2) $ supersymmetry. The supersymmetry  breaking preserves $R$-symmetry when the two gravitino masses are equal.  The set up is then a Dirac gravitino made of two gravitinos $\psi^{\mu}_{3/2, 1}$ and $\psi^{\mu}_{3/2, 2}$ with the same mass $m_{3/2}$, each of them coupling to a different sector with $N=1_1$ for the first and $N=1_2$ for the second. The coupling strength is in both cases given by $\frac{1}{m_{3/2} M_{Pl}}$, but may couple different gravitinos to different sectors. We will not present here an explicit realisation as it goes beyond the scope of this paper.

The second issue is very model dependent. A gravitino LSP can be produced in two different ways: either through thermal production or through decays of unstable sparticles. The relic density is expected to receive contributions from the two processes by an amount that depends on the particular spectrum and in the specific thermal history of the universe (for instance depending on the value of the reheat temperature), the possible decays of the inflatons, moduli, etc. The addition of a second gravitino might help the gravitino LSP to be a dark matter candidate by increasing the number of relativistic degrees of freedom at the freezout temperature. However, with  a second sector, it will make the story even more dependent of the details. We will not pursue this scenario further here.

The neutralino sector that we shall consider differs from that of the MSSM in two aspects:
\begin{itemize}
\item The LSP is now a linear combination of six states. There are now two additional states compared to the MSSM case.

\item In addition to the MSSM parameters, we have six additional parameters:
\begin{equation}
M'_1, M'_2, m_{1D}, m_{2D},  \lambda_S\ {\rm and} \;\lambda_T
\end{equation}
\end{itemize}

In this work, our purpose is to find a region of these parameters where the relic density is compatible with the dark matter made of a thermally produced neutralino LSP and compare this with the case of MSSM. We will assume both the $\mathbf{S}$ and $\mathbf{T}$ scalar are very  heavy and decoupled from the thermal bath.

\subsection{The Dirac case}
The first simplest case we will consider is the LSP to be mainly bino like.

\begin{equation}
\left(\begin{array}{c c c c c c}
0  & m_{1D} & 0     & 0     &  \frac{ \sqrt{2} \lambda_S }{g'}m_Z s_W s_\beta &    \frac{ \sqrt{2} \lambda_S }{g'}m_Z s_W c_\beta  \\
m_{1D} & 0  & 0     & 0     & -m_Z s_W c_\beta &   m_Z s_W s_\beta  \\
0     & 0     & 0  & m_{2D} & - \frac{ \sqrt{2} \lambda_T  }{g}m_Z c_W s_\beta & - \frac{ \sqrt{2} \lambda_T  }{g}m_Z c_W c_\beta  \\
0     & 0     & m_{2D} & 0   &  m_Z c_W c_\beta & - m_Z c_W s_\beta  \\
\frac{ \sqrt{2} \lambda_S }{g'}m_Z s_W s_\beta & -m_Z s_W c_\beta & -\frac{ \sqrt{2} \lambda_T  }{g}m_Z c_W s_\beta &  m_Z c_W c_\beta & 0    & -\mu \\
\frac{ \sqrt{2} \lambda_S }{g'}m_Z s_W c_\beta &  m_Z s_W s_\beta & -\frac{ \sqrt{2} \lambda_T  }{g}m_Z c_W c_\beta & -m_Z c_W s_\beta & -\mu & 0    \\
\end{array}\right) 
\label{diracgauginos_NeutralinoMassarraycase1}
\end{equation}
With $\mu > m_{1D}$ The DG-adjoint vevs are now
\begin{eqnarray}
v_s &\simeq &\frac{v^2}{2(M_S^2+m_S^2+4 m^2_{1D}+ B_S)} \ \ { \left[g'    m_{1D} c_{2\beta} + {\sqrt{2}} \mu    \lambda_S     \right]} \nonumber \\
v_t & \simeq &  \frac{v^2}{2(M'2+m_T^2+4 m^2_{2D}+ B_T)} \ \ \left[ g    m_{2D}  c_{2\beta} -{\sqrt{2}} \mu    \lambda_T + \frac{\lambda_T }{\sqrt{2}} (  M_T+    A_T)  s_{2\beta}  \right].
\end{eqnarray}

To have a pure Dirac LSP, the lightest two eigenvalues of the neutralino mass matrix \ref{diracgauginos_NeutralinoMassarraycase1}  
most form a pair of equal magnitude but opposite sign. It is straightforward to show that in the case that both $\lambda_s, \lambda_T$ take their $N=2$ values we will only have Dirac neutralinos. To do this, we note that solving for the eigenvalues of the neutralino mass matrix $\mathcal{M}$ means solving the equation $f(\lambda) \equiv \det(\mathcal{M}-\lambda)=0$. For purely Dirac states, $f(\lambda) = \prod_{i=1}^3 (\lambda^2 - a_i^2)$ for eigenvalues $\pm a_i$, which is true if and only if $f(-\lambda) = f(\lambda)$. By examining the coefficients of $\lambda^1$ and $\lambda^3$ (
 that of $\lambda^5$ being automatically zero due to the tracelessness of $\mathcal{M}$) we find that this requires
\begin{eqnarray}
\lambda^1 &:& 2\mu c_\beta s_\beta M_Z^2 \bigg[ c_W^2 m_{1D}^2 \bigg( 1 - \frac{2\lambda_T}{g^2} \bigg) + s_W^2 m_{2D}^2 \bigg( 1 - \frac{2\lambda_S}{(g')^2} \bigg)\bigg] = 0 \nonumber \\
\lambda^3 &:& 2\mu c_\beta s_\beta M_Z^2 \bigg[ c_W^2  \bigg( 1 - \frac{2\lambda_T}{g^2} \bigg) + s_W^2 \bigg( 1 - \frac{2\lambda_S}{(g')^2} \bigg)\bigg] = 0
\end{eqnarray}
proving the above assertion. 

If we assume a mostly bino/$U(1)$ adjoint LSP, by assuming that $m_{1D} < m_{2D} \ll \mu$ we can expand the LSP eigenvalues to next to leading order in $m_{1D}^2/\mu, m_{2D}^2/\mu$ to find
\begin{equation}
m_{LSP} = m_{1D} + \frac{M_Z^2 s_W^2}{(g^{\prime})^2\mu}\bigg[ \sqrt{2}\lambda_S g' (s_\beta^2 - c_\beta^2) \pm (2\lambda_S^2 - (g')^2 )c_\beta s_\beta  \bigg] + ...
\label{BinoLSPmass}\end{equation}
Assuming that $(2\lambda_S^2 - (g')^2 )c_\beta s_\beta$ is negative, the LSP takes the upper sign, and the mass splitting is given to lowest order by
\begin{equation}
\Delta m_{LSP} = -2 \frac{M_Z^2 s_W^2}{\mu} \frac{(2\lambda_S^2 - (g')^2 )}{(g^{\prime})^2}c_\beta s_\beta.
\end{equation}
Note that this reduces to the result of \cite{Hsieh:2007wq} when $\lambda_S \rightarrow 0$. The eigenstates at this order can be read off from the rotation matrices, given in appendix \ref{Appendix:BinoWino}.

For the case of a Higgsino LSP, where $m_{1D} \sim m_{2D} \gg \mu$, we find
\begin{equation}\begin{split}
m_{LSP} =&\mu + \frac{\sqrt{2} (s_\beta^2 - c_\beta^2)M_Z^2}{g g' m_{1D} m_{2D}} \bigg[ c_W^2 g' \lambda_T m_{1D}  + s_W^2 g \lambda_S m_{2D} \bigg]\\
& + \frac{M_Z^2}{\mu} \bigg[ \frac{4 c_\beta^2 s_\beta^2 M_Z^2 (c_W^2 g' \lambda_T m_{1D} + g \lambda_S m_{2D} s_W^2)^2}{g^2 (g')^2 m_{1D}^2 m_{2D}^2 }\bigg] \\
&- \mu  \bigg[ \frac{M_Z^2 [c_W^2 (g')^2 (g^2 + 2 \lambda_T^2) m_{1D}^2 +  g^2 ((g')^2 + 2 \lambda_S^2) m_{2D}^2 s_W^2] }{2 g^2 (g')^2 m_{1D}^2 m_{2D}^2 }\bigg]\\
&\mp \frac{c_\beta s_\beta \mu M_Z^2}{g^2 (g')^2 m_{1D}^2 m_{2D}^2 } \bigg[  (g')^2 m_{1D}^2 c_W^2 (g^2 - 2 \lambda_T^2) + g^2 m_{2D}^2 s_W^2 ((g')^2 - 2 \lambda_S^2)  \bigg]
\end{split}\end{equation}
where we have had to expand to the second order to obtain a mass splitting.

If we suppose that the adjoint couplings take the $N=2$ values at some scale $M_{N=2}$ and we run $\lambda_S, \lambda_T$ down to the supersymmetry breaking scale $M_{N=1}$ (equal to $m_S$ or $m_T$) then we can generate a mass splitting since the adjoints do not couple to the matter multiplets, and the Higgs' wavefunction renormalisation also contributes. To leading order we have

\begin{eqnarray}
\bigg[2\lambda_S^2  - (g')^2\bigg]_{M_{N=1}} &=& - \frac{2(g')^2}{16\pi^2}  \bigg[ 3 |y_t|^2 + 3 |y_b|^2 + |y_{\tau}|^2 -10 (g')^2\bigg] \log (\frac{M_{N=2}}{M_{N=1}}) \nonumber \\
\bigg[2\lambda_T^2  - g^2\bigg]_{M_{N=1}} &=& - \frac{2g^2}{16\pi^2}  \bigg[ 3 |y_t|^2 + 3 |y_b|^2 + |y_{\tau}|^2 - 4 g^2 \bigg] \log (\frac{M_{N=2}}{M_{N=1}})
\end{eqnarray}
If we assume that $y_t \sim 1$ and the other couplings much smaller, then we find for a Bino LSP that the mass difference should be
\begin{eqnarray}
\Delta m_{LSP} &\approx& 2 c_\beta s_\beta \frac{M_Z^2 s_W^2}{\mu} \frac{3}{8\pi^2} \log (\frac{M_{N=2}}{M_{N=1}}) \nonumber\\
&\approx& 0.15 {\rm GeV} \ \ (\frac{TeV}{\mu})\ \  \frac{t_\beta}{1+t_\beta^2} \log (\frac{M_{N=2}}{M_{N=1}}).
\label{Neq2MassSplit}
\end{eqnarray}
where, obviously, smaller splitting can be seen to correspond to larger values of $\tan{\beta}$ and smaller ratio ${M_{N=2}}/{M_{N=1}}$. For instance, taking $\mu= 1$TeV, $M_{N=2}\sim 10^{16}$GeV, and $t_\beta = 50$ leads to 
\begin{equation}
\Delta m_{LSP} \approx  4.4 \frac{t_\beta}{1+t_\beta^2} \ \  {\rm GeV} \approx 90 {\rm MeV}
\label{Neq2MassSplitex1}
\end{equation}
while for $\mu= 1$TeV, $M_{N=2}\sim 10^{4}$GeV, and $t_\beta = 50$, we obtain 
\begin{equation}
\Delta m_{LSP} \approx  0.3 \frac{t_\beta}{1+t_\beta^2}  \ \ {\rm GeV} \approx 6 {\rm MeV}.
\label{Neq2MassSplitex2}
\end{equation}
The latter case could be realized in low cut-off scale  models (such as for large extra-dimensions).

\subsection{Dark matter relic abundance}

In the DG model, as in the MSSM, there are several scenarios that lead to a relic abundance of the neutralino LSP, 
$\Omega h^2 \approx  0.11$. These include naturally  the MSSM-like scenarios   
\begin{itemize} 
\item{}a mixed bino/Higgsino LSP that annihilates mainly into W pairs (or top pairs)
\item{}a mixed bino/wino or bino/wino/Higgsino that annnihilates mainly into W pairs
\item{}a bino that annihilates into fermion pairs when sleptons are light, a significant region of parameter space
where this process occurs has been ruled out by LEP. 
\item{}a bino that coannihilates with sfermions. 
\item{}an almost pure bino with mass $2 m_{\bino} \approx  m_{h,A}$ with efficient 
annihilation through Higgs exchange.
\end{itemize}
New dark matter scenarios occur as well. Firstly in the special case of a pure Dirac bino LSP 
(or an almost pure Dirac bino), annihilation into light fermion pairs becomes  efficient. This is because the
process does not have a strong p-wave suppression as in the Majorana neutralino case~\cite{Hsieh:2007wq}. 
Secondly the MSSM-like scenarios where  $\bino,\wino$ are replaced with $\binop$ and $\winop$ can also occur.
These are found either with Dirac masses or with Majorana masses.
Typically, as we will see in case studies below,  these involve more coannihilation processes.

In the DG model, the dark matter detection properties can  be quite different than in the usual MSSM. 
For one, the annihilation of a Dirac neutralino into light fermion pairs is not suppressed, even at $v/c\approx 0.001$. 
More generally, the content of the LSP that determines the coupling of the LSP to other particles can be different in
the DG model. This is largely due to the additional $\binop$ or $\winop$ components. 
For example, the spin independent elastic scattering rate on nucleons that is  dominated by Higgs exchange
unless squarks are light, depends on the higgsino content of the LSP. In the DG model, 
the higgsino fraction is often suppressed leading to small rates.

In the following subsections we present results for a few case studies.
For each scenario, we find the parameter space that predicts a relic abundance compatible with the value
 determined by  cosmological measurements, $\Omega h^2=0.113\pm 0.0034$~\cite{Komatsu:2008hk} and examine the predictions for the direct and indirect
detection rates.  In all cases special attention is paid to the typical DM scenarios with either a 
$\bino(\binop)$ or mixed $\bino(\binop)/\tilde{h}$  or $\bino(\binop)/\wino(\winop)$ scenarios (here $\tilde{h}$ stands for $\tilde{H_u^0}$
or $\tilde{H_d^0}$).
The  possibility of efficient annihilation through a Higgs resonance
is a generic feature of all models where s-channel resonance can occur, 
we will however  not consider this mechanism in detail
as it requires fine-tuning of the model parameters.

\section{Results}

All numerical results are based on micrOMEGAs2.3 for the computation of the spectrum, the relic abundance, the elastic scattering rate 
as well as the annihilation rate $\sigma v|_0$ relevant for indirect detection~\cite{Belanger:2006is, Belanger:2008sj}. We have implemented in this code the DG model described in section 3. 
The one-loop quark/squark corrections to the Higgs masses are computed 
as well as the $\lambda_S,\lambda_T$ corrections to the
effective potential~(\ref{potential4}). The latter can increase the Higgs mass by a few GeV as compared with the MSSM. 
We have imposed the LEP bounds on charged sparticles as well as on the Higgs mass ($m_h>111$~GeV), 
allowing a large theoretical uncertainty for $m_h$ since two-loops corrections are ignored. In all cases we
fix $\tan\beta=10$, $A_t=-1.5$TeV and $M_{\tilde{q}}=1$TeV. Under these conditions the Higgs mass limit is easily
satisfied. The elastic scattering rates are computed taking the micrOMEGAs default values for the quark coefficients in the
nucleon~\cite{Belanger:2008sj}. Varying these coefficients could induce large corrections to the predicted rate.

\subsection{Pure Dirac masses : $M_1=M'_1=M_2=M'_2=0$}

In the case of pure Dirac masses $M_{1D},M_{2D}\neq 0$, 
the LSP can be a Dirac fermion provided $\lambda_S$ and $\lambda_T$ take their N=2
value, eq.~(\ref{Neq2Lambdas}).  In general  one expects a mass splitting generated by the N=2
breaking effect, eq.~(\ref{Neq2MassSplit}),  nevertheless it is possible to tune the parameters of the model  such that the two
lightest Majorana states are degenerate and  make a Dirac fermion. The
implications for dark matter detection of a Dirac fermion are important so this case is worth
consideration.

\subsubsection{Dirac fermion}

\begin{figure}
\setlength{\unitlength}{1mm} \centerline{\epsfig{file=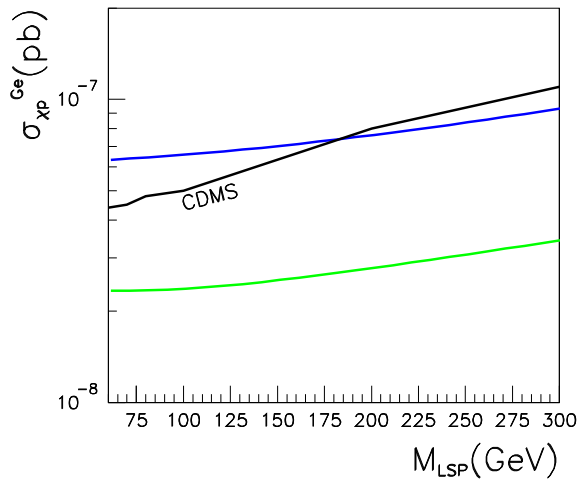,width=9cm}}
\vspace{-1cm}
\caption{a)  Effective neutralino-nucleon elastic scattering
cross section  as a function of $\mneut$ for a Dirac LSP case with $m_{\tilde q}=1$~TeV
(blue) and $m_{\tilde q}=1.2$~TeV(green). For each LSP mass, the
common right-handed slepton mass is adjusted so that $\Omega h^2=0.11$. The CDMS limit is
also displayed (black).}
\label{fig:dirac}
\end{figure}

A Dirac neutralino, contrary to a Majorana neutralino,  
can annihilate into light fermion pairs with a large rate,
 thus offering an explanation to the  excess of positrons seen by Pamela~\cite{Adriani:2008zr,Harnik:2008uu} without spoiling the
 antiproton observations~\cite{Adriani:2008zq} {\footnote{Note that the recent Fermi results~\cite{Abdo:2009zk} on the total electron and positron spectrum
 is in agreement with the positron spectrum measured by PAMELA.}}.
In particular the Dirac bino can have a  much larger annihilation rate into leptons than into quarks
when the mass of the right-handed sleptons are of the order of the bino mass. 
 However, the Dirac neutralino has
an effective vectorial interaction with quarks in the nucleon, this leads to
potentially large rates for direct detection. For a Dirac fermion the spin-independent 
elastic scattering cross section
receives dominant contributions from squark and Z exchange. To avoid exceeding the experimental bound~\cite{Ahmed:2008eu,Angle:2008we}
it is enough to 1) fix the mass of the  squark 
that couples most strongly to the bino, the one with the largest hypercharge, to $m_{\tilde{u}_R} > 1.- 1.2$~TeV 
and 2) suppress the higgsino LSP component such that the coupling of the LSP to the Z is reduced. 
To illustrate this we have computed the neutralino nucleon  elastic scattering cross section in a scenario where 
$M_{2D}=1.5 M_{1D}$, $\mu=1$~TeV, $M_{{\tilde f}_L}$~TeV and $M_{{\tilde q}_R}=1 {\rm or} 1.2$~TeV. 
The value of the common mass for the right-handed sleptons, $M_{\tilde{l}_R}$, is adjusted for each value of $M_{1D}$ 
such that  $\Omega h^2=0.11$.  The lower value of the squark mass leads to a cross-section exceeding the CDMS limit for light neutralinos, 
see Fig.~\ref{fig:dirac}. Note that
one characteristic of  scenarios with a Dirac particle as DM is that the spin independent amplitude for elastic scattering on nucleons
can be different  for protons and neutrons, for example $\sigma_{\neut p}<<\sigma_{\neut n}$ 
if Z exchange dominates. However the  experimental limits  on $\sigma_{\neut p}$ 
are extracted the amplitudes for protons and neutrons to be equal. 
In order to be able to compare directly with the experimental limit from CDMS~\cite{Ahmed:2008eu}, we rescale the nucleon cross section and 
define an effective  $\sigma^{Ge}_{\neut p}=(Z f_p+(A-Z) f_n)^2/A^2$ where $f_{p(n)}$ are the amplitudes on nucleons and $Z=32,A=76$ for Germanium.
In this scenario, the rate for indirect detection is large, 
$\sigma v|_{ll} \approx 2.5\times 10^{-26}~{\rm cm}^3/{\rm sec}$ for $M_{1D}=300$~GeV and
the lepton channels are almost two orders of magnitude larger than the quark channels.

For the remainder of this section we will consider the case where the LSP is a Majorana fermion,
and the LSP-NLSP mass splitting in the MeV to GeV range as illustrated by eq.~ (\ref{Neq2MassSplitex1}) and eq.~ (\ref{Neq2MassSplitex2}). This results
from shifts of the value of $\lambda_S$  (say by only 1\%) from the N=2 value in eq.~(\ref{Neq2Lambdas}). 
Because of the small mass splitting  between the LSP and the NLSP  the $\neut\neutt$  channels will not suffer
from a  Boltzman suppression and will  contribute significantly to the effective
annihilation cross section that enters the standard computation of the relic
abundance~\cite{Belanger:2006is}.

\subsubsection{Bino-LSP}

\begin{figure}
\setlength{\unitlength}{1mm} \centerline{\epsfig{file=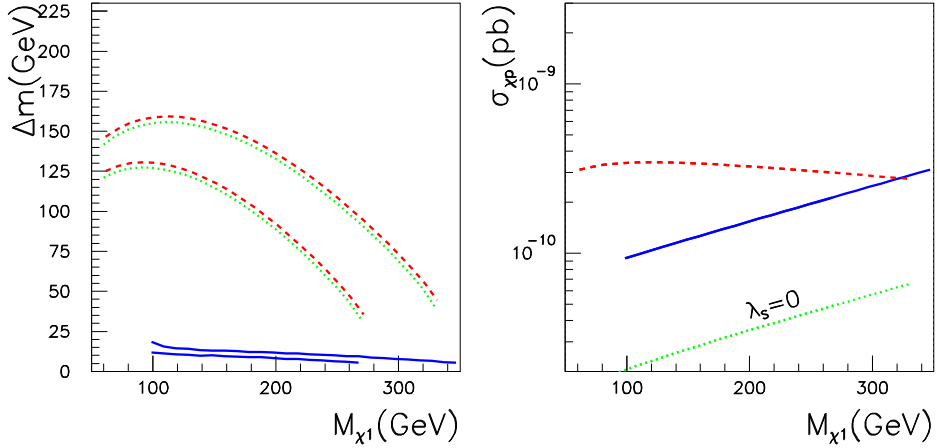,width=14cm}}
\caption{a)Contours of  $\Omega h^2=0.094$ (upper) and $\Omega h^2=0.094$ (lower) in the $\Delta(m_{\tilde\tau}-\mneut)$ vs $\mneut$ plane 
in the MSSM(full blue line), the DG model(red dashed line) and the
DG model with $\lambda_S=0$ (green dotted line). b) Elastic scattering  $\sip$ as a function of $\mneut$ - same colour code as a)}
\label{fig:sl}
\end{figure}

The main mechanism for annihilation of a bino LSP is through exchange of sfermions, the sfermions with 
largest hypercharge, the right-handed(RH) sleptons, giving the dominant contribution. 
In general this process  gives only $\Omega h^2 \approx {\cal O}(1)$. This is because 
$\sigma\propto\mneut^2/m_{\tilde f_R}^4$. 
and both the neutralino and the sfermion need to be near 100GeV, that is near  the LEP exclusion region,  
to reach $\Omega h^2=0.1$. Slepton coannihilation  provides an alternative for reducing the relic
abundance.  With Dirac mass terms and nearly degenerate $\bino$ and $\binop$, the processes
$\neut\neutt\rightarrow f\bar{f}$ provide the dominant annihilation mechanism 
~\cite{Hsieh:2007wq}. One can obtain $\Omega h^2\approx 0.1$ even with sleptons twice as heavy as   
the bino-LSP. This is displayed in  Fig.~\ref{fig:sl} where we compare the slepton-LSP mass splitting 
that produces a relic abundance in the  WMAP range in the DG model and in the MSSM. Here we show contours including a large theoretical
uncertainty in the determination of $\Omega h^2= 0.11\pm 0.026$.
  In both cases we fix $\mu=1$~TeV and take $M_{2D}=2M_{1D}$(DG model) or $M_2=2M_1$(MSSM), so that $\mneut\approx M_{1D}(M_1)$. 
  For each neutralino mass we vary the RH slepton masses to find the given relic abundance contour. 
  Because of the mixing in the stau sector, the stau turns out to be the lightest slepton. 
In this scenario the elastic scattering cross-section is small ($\sip\approx 10^{-10}$~pb) although
within the reach of detectors such as Xenon~\cite{Aprile:2009yh}. The small cross sections  can be linked to the LSP higgsino composition. 
The  indirect detection cross section is also small in both models $\sigma v|_0\approx 10^{-29} {\rm cm^3}{\rm sec}$.  

The prediction for the elastic scattering cross-section can be shifted significantly for a different choice of 
 $\lambda_S$. For example for $\lambda_S=0$, the higgsino fraction $f_H$ ($\equiv |N_{i5}|^2 + |N_{i6}|^2$, $i$ denoting the index of the LSP, in this case $1$) of the LSP 
  decreases (see appendix \ref{Appendix:BinoWino}) thus suppressing the elastic scattering rate further 
   by an order of magnitude. 
  Of course  $\lambda_S=0$  will increase the $\bino-\binop$ mass splitting, 
  yet a splitting of only  a few GeVs has little impact on the relic abundance(fig.~\ref{fig:sl}b).
Conversely enhanced rates can be  found for $\lambda_s>g'/\sqrt{2}$.

The bino LSP up to a few hundred GeV is therefore a natural DM candidate in the DG model. The discovery 
of such neutralinos and sleptons of a few hundred GeV  is within the reach of the LHC. Furthermore, only 
 a rough estimate of the slepton/LSP mass difference would be sufficient to point out an 
inconsistency between  the relic abundance of dark matter obtained from cosmological mesurements
and the one predicted from collider measurements of the SUSY spectrum if done  in the context of the MSSM.  

\subsubsection{Mixed bino/higgsino LSP}

A mixed bino/higgsino LSP is a 
more natural DM candidate than the bino  because of the efficient annihilation into gauge boson
pairs or top quark pairs. In particular the annihilation into gauge boson pairs proceeds through t-channel chargino exchange or Z/H exchange. 
For all annihilation diagrams some higgsino or $\wino,\winop$ fraction of the LSP is involved, see the explicit expressions for the LSP couplings in Appendix C. Here and in the following we assume heavy sleptons $M_{\tilde l_{L}}=M_{\tilde l_{R}}=1$~TeV as they are not needed for efficient annihilation. 

\begin{figure}
\setlength{\unitlength}{1mm} \centerline{\epsfig{file=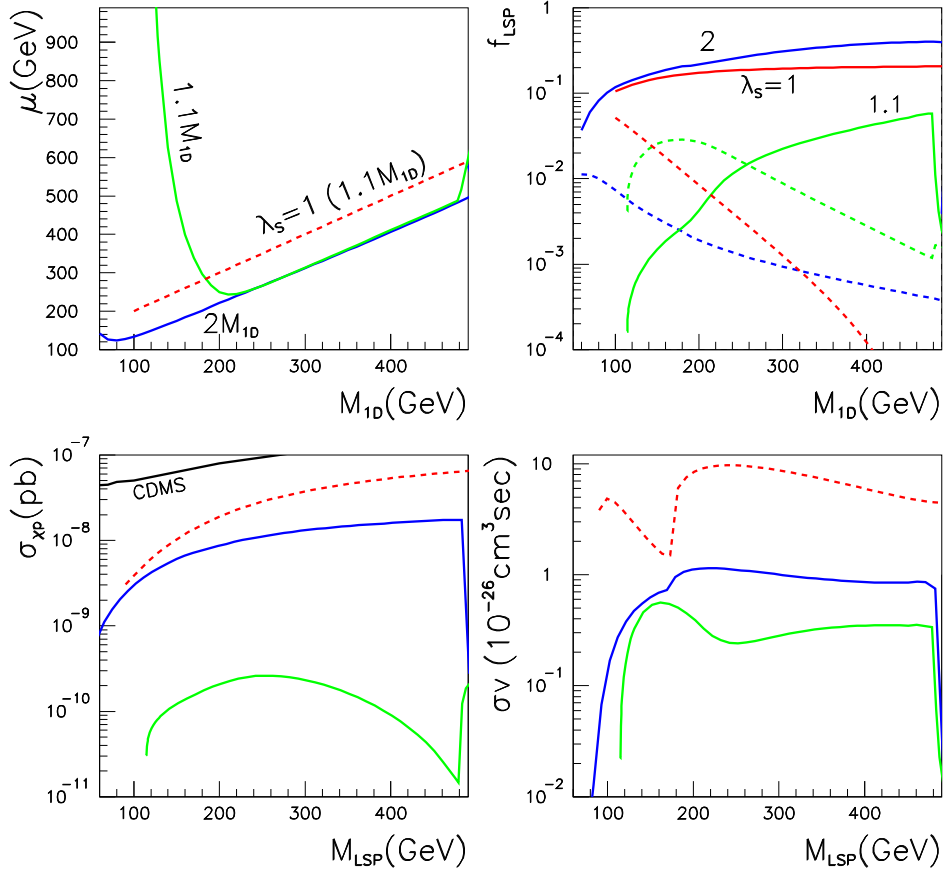,width=14cm}}
\caption{a) Contours of $\Omega h^2=0.11$ in the $\mu-M_{1D}$ plane in the DG model for the mixed bino/Higgsino $M_{2D}=2M_{1D}$ (full blue line) 
and for $M_{2D}=1.1M_{1D}$(full green line), $M_{2D}=1.1M_{1D}, \lambda_S=1$(red dashed line) b) Content of the LSP, $f_h$ (full lines) 
$f_W$ (dashed lines),  same colour code as a),
 c) $\sigma_{\neut p}$ as a function of $\mneut$ - same colour code as a) with the CDMS limit (full black line)
 d) $\sigma v|_0$ as a function of $\mneut$ - same colour code as a)}
\label{fig:hw_dg}
\end{figure}

\begin{figure}
\setlength{\unitlength}{1mm} \centerline{\epsfig{file=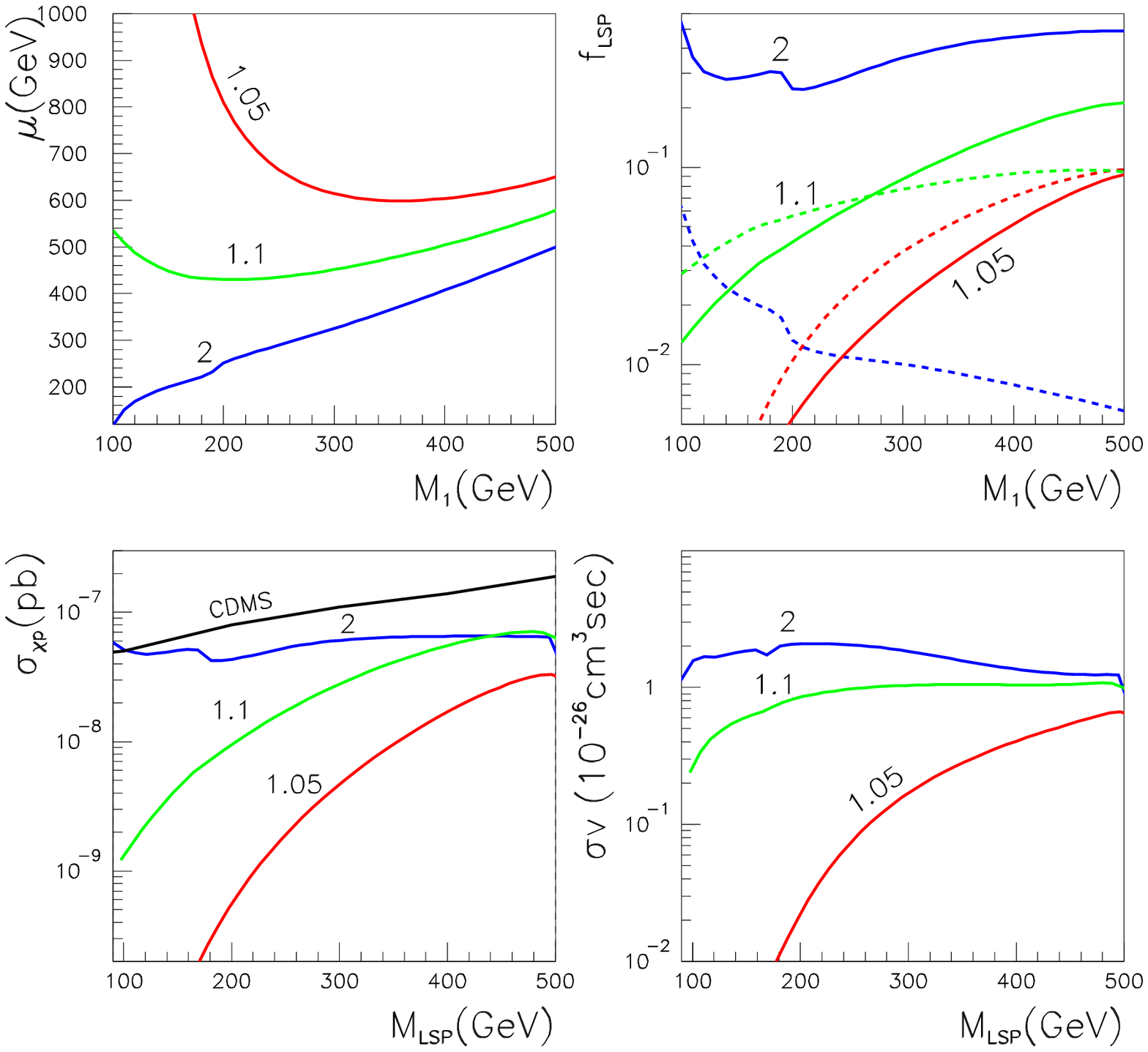,width=14cm}}
\caption{a) Contours of $\Omega h^2=0.11$  in the  $\mu- M_1$ plane in the MSSM for the mixed bino/higgsino, $M_{2}=2M_{1}$ (blue) 
and for $M_{2}=1.1M_{1}$ (green) and $M_{2}=1.05 M_{1}$ (red) b) Content of the LSP, $f_h$ (full lines) $f_W$ (dashed lines) - same colour code as a)  
 c) $\sigma_{\chi p}$ as a function of $\mneut$ - same colour code as a)  with the CDMS limit (full black line)
 d) $\sigma v|_0$ as a function of $\mneut$ - same colour code as a)}
\label{fig:hw_mssm}
\end{figure}

As a sample scenario we take  $M_{2D}=2 M_{1D}$ so that the wino fraction of the LSP is small and
fix the mass of all sfermions to 1 TeV. In the DG model we find the contour of $\Omega h^2=0.11$ in the $\mu-M_{1D}$ plane, 
see  fig.~\ref{fig:hw_dg}. The contour corresponds to  $\mu \approx M_{1D}$ and the region below the contour gives $\Omega h^2 < 0.11$. 
The LSP is dominantly bino/bino' with a higgsino fraction that  ranges  from  2-30\% 
along this contour as one increases the LSP mass. The small higgsino fraction of the LSP implies a small annihilation cross section into
W pairs. This  is however compensated by the contribution of the coannihilation channels to the relic abundance,  with   
$\tilde{\chi}_1^0 \tilde{\chi}_2^0, \tilde{\chi}_1^0 \tilde{\chi}^+, \tilde{\chi}_2^0 \tilde{\chi}_2^0, \tilde{\chi}_1^0 \tilde{\chi}_3^0$ into gauge bosons or heavy quarks.
This is to be contrasted with the MSSM, where a similar contour in the $\mu-M_1$ plane is found but which feature a Higgsino fraction  $f_H\approx 30\%$. 
Since coannihilation channels do not enter  direct/indirect detection rates, 
the rates are suppressed in the 
DG model relative to the MSSM, compare fig.~\ref{fig:hw_dg} with fig.~\ref{fig:hw_mssm}. The sharp variation in $\sigma_{\neut p}$ at the edge of the figure 
correponds to the onset of the efficient annihilation of neutralinos through a Higgs resonance. 
Note that the mass spectra in the two models are rather similar, apart from the fact that there are two
nearly degenerate  $\bino,\binop$  and an additional chargino in the DG model.

\subsubsection{Mixed bino/wino/higgsino LSP}

We consider finally the case of a mixed bino/wino LSP, for example we choose  $M_{2D}=1.1 M_{1D}$.
This choice for the  gaugino masses  necessarily implies a roughly 10\% mass splitting between the LSP, an 
equal mixture of $\bino/\binop$, and  
the $\neutt$($\bino/\binop$),$\tilde{\chi}_3^0,\neu_4(\wino,\winop)$ and  the $\charg,\chargt$. 
Thus naturally one finds important contributions from a variety of coannihilation channels. 
The contour of $\Omega h^2=0.11$ is displayed in fig.~\ref{fig:hw_dg}.  
Along this contour the coannihilation into gauge boson pairs or fermions dominate. 
Note that the  main difference with the bino/higgsino case occurs for a light LSP, in particular, 
 the small wino fraction is sufficient to provide efficient annihilation through chargino exchange,
thus larger values of $\mu$ are allowed. On the other hand in an MSSM model with a mixed wino LSP, here we  
choose $M_2=1.1 M_1$, only about half the effective annihilation cross section comes from coannihilation with 
$\charg,\neutt$, the dominant mode is LSP annihilation into W pairs (or top pairs). A smaller gaugino mass splitting, for example  
$M_2=1.05 M_1$ requires an even smaller higgsino fraction.

The direct detection rate which relies mainly on the LSP higgsino fraction is lower than for the mixed Higgsino LSP, this is 
especially true for light LSPs, see fig.~\ref{fig:hw_dg}c.
As in other scenarios, the elastic scattering rate in the DG model is suppressed as compared to the MSSM. This statement is 
strongly dependent on the value of $\lambda_S$. For example for  $\lambda_S=1$ we found a rate that increases by one or two orders of magnitude,
this means within the range of the next run of Xenon~\cite{Aprile:2009yh}. 
The indirect rate is also increased in this case.

\subsection{Large Majorana masses $M_1=M_2=1$~TeV} 

\begin{figure}
\setlength{\unitlength}{1mm} \centerline{\epsfig{file=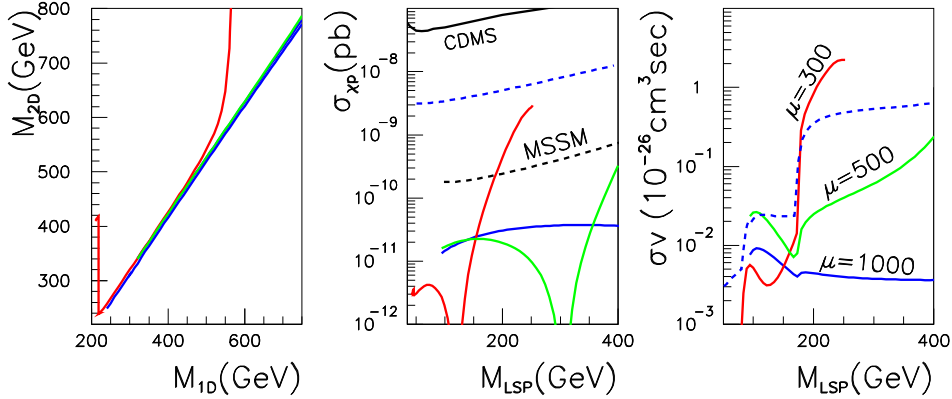,width=14cm}}
\caption{a) Contours of $\Omega h^2=0.11$  in the  $M_{2D}-M_{1D}$ plane in the DG model for $\mu=1000$~GeV(blue), $\mu=500$~GeV(green) and
$\mu=300$~GeV(red)
 b) $\sigma_{\chi p}$ as a function of $\mneut$ - same colour code as a). The dash curves correspond to $\mu=1000$~GeV (MSSM) and
 $\lambda_S=1$ (DG model)
 c) $\sigma v|_0$ in the DG model as a function of $\mneut$ - same colour code as b).}
\label{fig:maj}
\end{figure}

First consider the  $M_1'=M_2'=0$ case. When $\mu=1$~TeV, the
LSP is dominantly $B'$ with some $B$ admixture and with  a mass $m_{LSP}\approx m_{1D}^2/M_1$.
Both the wino and higgsino fraction of the LSP are very small, therefore the usually efficient
annihilation channel into W pairs is suppressed. A value of $\Omega h^2=0.11$ can only be
reached  because of the  coannihilation channels such as $\neutt
\charg,\charg\chargm$ into gauge boson pairs. These channels proceed through the wino
component of the heavier neutralinos and charginos. For coannihilation to work one needs roughly
a $10\%$ mass difference between the LSP and the neutral and charged NLSP, NNLSP. 
In that sense the weak scale chargino/neutralino sector is similar to the one of the bino/wino scenarios in the MSSM, 
since additional states are at the TeV scale.  
In  fig.~\ref{fig:maj} we display the contour $\Omega h^2=0.11$ in the $m_{2D}$-$m_{1D}$ plane for different values of
$\mu$. When  the higgsino fraction of the LSP  increases, more precisely 
when the LSP mass becomes comparable to $\mu$, annihilation becomes very efficient and  
the relic abundance is always $\Omega h^2<0.11$.  For  $\mu=300$~GeV,  this occurs for $M_{1D}\approx 550$~GeV, or 
$\mneut\approx 260$~GeV.

As discussed in previous scenarios,  the predictions of the DG model for elastic scattering cross sections  are usually
suppressed when compared to an equivalent MSSM scenario. This is related to the fact that the LSP has a lower Higgsino
fraction and that the relic abundance relies more heavily  on coannihilation in the DG model. For example for 
$\mu=1000$~GeV the rate is suppressed by one order of magnitude in the DG model,
see fig.~\ref{fig:maj}b. Here the MSSM rate corresponds to the $\Omega h^2=0.11$
contour for $\mu=1000$~GeV. As before, a large increase in the elastic scattering rate is expected 
when the LSP has a significant higgsino
fraction. This occurs when $M_{LSP}\approx \mu$ or when $\lambda_S\neq g'/\sqrt{2}$, see for example the contour $\mu=1$~TeV,
$\lambda_S=1$ in fig.~\ref{fig:maj}b. Note that when the detection rate is small, interference between the squark and Higgs exchange can lead to a further suppression of the detection rate, see the dips for the $\mu=300,500$~GeV scenarios in  fig.~\ref{fig:maj}b.

The self-annihilation of the LSP at $v=0$ is small for the $\bino(\binop)/\wino$ scenario, 
the domimant channels are $W^+W^-$ or $t\bar{t}$ when this channel becomes kinematically accessible. As usual, the rate increases significantly 
with the higgsino content of the LSP, see fig.~\ref{fig:maj}c.

We will not discuss the more general case of the DG model where all Majorana and Dirac mass terms are
present. When these masses are at the electroweak scale, there is no difficulty in finding  dark matter
scenarios with $\Omega h^2\approx0.1$ and a LSP below the TeV scale.  The LSP can be either a $\bino$ or mixed 
$\tilde B/\tilde h,\tilde B'/\tilde h, \tilde B/\tilde W(\tilde W')$. 
All neutralinos and charginos could be within the kinematical reach of the LHC, 
the discovery of additional neutralino/chargino states would be the most obvious way to distinguish this model 
from the MSSM.

\section{Conclusion}

Dirac masses for gauginos can be obtained by pairing the MSSM gauginos with additional singlet, triplet and octet states in the adjoint representation.
The model then contains additional neutralinos and charginos as well as new scalar particles. We have discussed both the case where the adjoint scalars decouple leaving a new quartic Higgs coupling in the effective potential and the one where the singlet remains light.
This model has R-parity so the LSP is stable,  the LSP could be a gravitino or a neutralino. 

We have made a first exploration of the parameter space of the model in the case of the neutralino LSP to find regions where the dark matter relic abundance is in agreement with cosmological measurements (assuming the standard cosmological scenario). Among the possible scenarios the ones that have a feature that distinguish them from the MSSM include the Dirac gaugino LSP that has a non suppressed annihilation into light fermions, the pseudo-Dirac bino LSP that (co-)annihilates into leptons via slepton exchange with sleptons much heavier than expected in the MSSM, as well as several mixed bino/wino/higgsino or bino/higgsino scenarios.
For the latter scenarios the direct/indirect  detection rates are often expected to be lower than in the MSSM, yet could be within the range of future detectors. In our numerical analysis we have concentrated on the DM candidates in the 100-500~GeV range, thus with several states within the range of the LHC. We have  avoided a detailed discussion of scenarios where the annihilation of the LSP is made efficient by the presence of a Higgs resonance, as these scenarios require fine-tuning the masses of the LSP and Higgses. We expect that additional annihilation channels involving the new scalars in the model with a weak scale extra singlet would  give new regions of parameter space with efficient DM annihilation as is found in the NMSSM. 
Finally we mention that the mass splitting between the two lightest neutralino states can be small although unless $\lambda_S$ is fixed exacty to its N=2 value the splitting is expected to be larger than the 100keV needed for inelastic dark matter scattering.

\section*{Acknowledgements}
K.B. and M.G. wish to thank P. Slavich for discussions. 
G.B. and A.P.  thank A. Semenov for his help with LanHEP. We also thank Jan Kalinowski for pointing out 
 to us a typo in the first version.
This work was supported in part by the GDRI-ACPP of CNRS. 
The work of A.P. was supported by the Russian foundation for Basic Research, 
grants RFBR-08-02-00856-a and RFBR-08-02-92499.

\appendix

\section{Rotation Matrices and Eigenvalues for Bino/Wino Neutralinos}
\label{Appendix:BinoWino}

In this appendix we give the first order rotation matrices for the neutralinos in the approximation that $\mu$ is much larger than the other masses. Recall that the Lagrangian contains a term
\begin{equation}
\mathcal{L} \supset - \frac{1}{2} \chi_i \mathcal{M}_{ij} \chi_j = - \frac{1}{2}  \chi_i^\prime \mathcal{M}^{diag}_{ij} \chi_j^\prime 
\end{equation}
we write 
\begin{equation}
\chi_i^{\prime} = \delta_{ii} R_{ij} \chi_j \equiv N_{ij} \chi_j 
\end{equation}
where $R_{ij}$ is a real orthogonal matrix and $\delta_{ii}$ a unitary diagonal matrix of phases that ensures that all of the masses are positive. For the case of that the neutralino is mostly bino, we assume that $m_{1D}, m_{2D}, M_1, M_1^\prime, M_2, M_2^{\prime}$ are of the same order and much smaller than $\mu$. Then the eigenstates prior to mixing have mass eigenvalues
\begin{equation}
m_1^\pm = \frac{1}{2} \bigg[(M_1 + M_1^\prime) \pm \sqrt{(M_1 - M_1^\prime)^2 + 4 m_{1D}^2} \bigg]
\end{equation}
and similarly for $m_2^\pm$. We then define 
\begin{eqnarray}
c_1 \equiv \cos \theta_1 \equiv \frac{ m_{1D}}{\sqrt{(M_1^\prime - m_1^+)^2 + m_{1D}^2}} \nonumber \\
s_1 \equiv \sin \theta_1 \equiv \frac{  m_1^+ - M_1^{\prime}}{\sqrt{(M_1^\prime - m_1^+)^2 + m_{1D}^2}} \nonumber \\
\end{eqnarray}
so that
\begin{equation}
\left(\begin{array}{cc} c_1 & s_1 \\ -s_1 & c_1 \end{array} \right) \left(\begin{array}{cc} M_1^{\prime} & m_{1D} \\ m_{1D} & M_1 \end{array}\right)\left(\begin{array}{cc} c_1 & -s_1 \\ s_1 & c_1 \end{array} \right) = \left(\begin{array}{cc} m_1^+ & 0 \\ 0 & m_1^- \end{array}\right)
\end{equation}
and again similarly for the Wino states. Then $\delta_{11} = \sqrt{m_1^+/|m_1^+|},\delta_{22} = \sqrt{m_1^-/|m_1^-|}$ and similarly for the $m_2^\pm$ values. The matrix $R_{ij}$ is given by 
\begin{eqnarray}
\!\!\!\!\!R_{11}\!\!\!\!\!&=&\!\!\!\!\!c_1 - \frac{M_Z^2 s_1 s_W^2  (2 c_1 c_\beta  s_1 s_\beta ((g')^2 + 2 \lambda_S^2) +\sqrt{2} g' (c_1^2-s_1^2) (c_\beta^2-s_\beta^2)  \lambda_S) }{(g')^2 (m_1^--m_1^+) \mu} \nonumber \\
\!\!\!\!\!R_{12}\!\!\!\!\!&=&\!\!\!\!\!s_1 + \frac{M_Z^2 c_1 s_W^2  (2 c_1 c_\beta  s_1 s_\beta ((g')^2 + 2 \lambda_S^2) +\sqrt{2} g' (c_1^2-s_1^2) (c_\beta^2-s_\beta^2)  \lambda_S) }{(g')^2 (m_1^--m_1^+) \mu} \nonumber \\
\!\!\!\!\!R_{13}\!\!\!\!\!&=&\!\!\!\!\!\frac{c_W s_W M_Z^2}{g g' (m_1^+ - m_2^-) (m_1^+-m_2^+)} \bigg[ (m_1^+- c_2^2 m_2^- - s_2^2 m_2^+ ) (\sqrt{2} c_\beta^2 (g^\prime) s_1-\sqrt{2} (g^\prime) s_1 s_\beta^2-4 c_1 c_\beta s_\beta \lambda_S) \lambda_T \nonumber \\
\!\!\!\!\!&& - g (m_2^--m_2^+) c_2 s_2 (2 c_\beta g' s_1 s_\beta + \sqrt{2} c_1 c_\beta^2 \lambda_S -\sqrt{2} c_1 s_\beta^2 \lambda_S) \bigg]\nonumber \\
\!\!\!\!\!R_{14}\!\!\!\!\!&=&\!\!\!\!\!\frac{c_W s_W M_Z^2}{g g' (m_1^+ - m_2^-) (m_1^+-m_2^+)} \bigg[ (m_1^+- c_2^2 m_2^+ - s_2^2 m_2^- ) (\sqrt{2} c_\beta^2 (g^\prime) s_1-\sqrt{2} (g^\prime) s_1 s_\beta^2-4 c_1 c_\beta s_\beta \lambda_S) \lambda_T  \nonumber \\
\!\!\!\!\!&&\!- \lambda_T(m_2^--m_2^+) c_2 s_2 (\sqrt{2} c_\beta^2 (g^\prime) s_1-\sqrt{2} (g^\prime) s_1 s_\beta^2-4 c_1 c_\beta s_\beta \lambda_S) )\bigg] \nonumber \\
\!\!\!\!\!R_{15}\!\!\!\!\!&=&\!\!\!\!\!\frac{M_Z s_W}{g' \mu} \bigg[ g' s_1 s_\beta + \sqrt{2} c_1 c_\beta \lambda_S \bigg] \nonumber \\
\!\!\!\!\!R_{16}\!\!\!\!\!&=&\!\!\!\!\!-\frac{M_Z s_W}{g' \mu} \bigg[ g' s_1 c_\beta - \sqrt{2} c_1 s_\beta \lambda_S \bigg] 
\end{eqnarray}
\begin{eqnarray}
\!\!\!\!\!R_{21}\!\!\!\!\!&=&\!\!\!\!\!-s_1 - \frac{MZ^2 c_1 s_W^2  (2 c_1 c_\beta  s_1 s_\beta ((g')^2 + 2 \lambda_S^2) +\sqrt{2} g' (c_1^2-s_1^2) (c_\beta^2-s_\beta^2)  \lambda_S) }{(g')^2 (m_1^--m_1^+) \mu}\nonumber \\
\!\!\!\!\!R_{22}\!\!\!\!\!&=&\!\!\!\!\!c_1 - \frac{MZ^2 s_1 s_W^2  (2 c_1 c_\beta  s_1 s_\beta ((g')^2 + 2 \lambda_S^2) +\sqrt{2} g' (c_1^2-s_1^2) (c_\beta^2-s_\beta^2)  \lambda_S) }{(g')^2 (m_1^--m_1^+) \mu}\nonumber \\
\!\!\!\!\!R_{23}\!\!\!\!\!&=&\!\!\!\!\!\frac{c_W s_W M_Z^2 }{ g (g^\prime) (m_1^--m_2^-) (m_1^--m_2^+) \mu} \bigg[ (m_1^- - m_2^+) (\sqrt{2} c_1 c_\beta^2 (g^\prime) \lambda_T - \sqrt{2} c_1 (g^\prime) s_\beta^2 \lambda_T + 4 c_\beta s_1 s_\beta \lambda_S \lambda_T)  \nonumber \\
\!\!\!\!\!&& + (m_2^- - m_2^+) (-2 c_1 c_2 c_\beta g (g^\prime) s_2 s_\beta +    4 c_2^2 c_\beta s_1 s_\beta \lambda_S \lambda_T  + \sqrt{2} (c_\beta^2 - s_\beta^2) ( c_2  g s_1 s_2 \lambda_S -   c_1 c_2^2 (g^\prime) \lambda_T)\bigg]\nonumber \\
\!\!\!\!\!R_{24}\!\!\!\!\!&=&\!\!\!\!\!\frac{c_W s_W M_Z^2 }{ g (g^\prime) (m_1^--m_2^-) (m_1^--m_2^+) \mu} \bigg[ (m_1^- - m_2^+)
 2 c_1 c_\beta g (g^\prime) s_\beta + \sqrt{2} g s_1 \lambda_S - 2 \sqrt{2} c_\beta^2 g s_1 \lambda_S)\nonumber \\
\!\!\!\!\!&& -( m_2^- -m_2^+) (\sqrt{2} (c_\beta^2 - s_\beta^2) (-g s_1 s_2^2 \lambda_S + c_1 c_2 (g^\prime) s_2 \lambda_T) + 2 c_\beta s_\beta (c_1 g (g^\prime) s_2^2 + 2 c_2 s_1 s_2 \lambda_S \lambda_T)\bigg] \nonumber \\
\!\!\!\!\!R_{25}\!\!\!\!\!&=&\!\!\!\!\!\frac{M_Z s_W}{g' \mu} \bigg[ g' c_1 s_\beta - \sqrt{2} s_1 c_\beta \lambda_S \bigg] \nonumber \\
\!\!\!\!\!R_{26}\!\!\!\!\!&=&\!\!\!\!\!-\frac{M_Z s_W}{g' \mu} \bigg[ g' c_1 c_\beta + \sqrt{2} s_1 s_\beta \lambda_S \bigg] 
\end{eqnarray}
\begin{eqnarray}
\!\!\!\!\!R_{31}\!\!\!\!\!&=&\!\!\!\!\!\frac{c_W s_W M_Z^2 }{ g (g^\prime) (m_1^+-m_2^+) (m_2^+- m_1^-) \mu} \bigg[ (c_1^2 m_1^-+ s_1^2 m_1^+ - m_2^+ ) \lambda_S (\sqrt{2} c_\beta^2 g s_2-\sqrt{2} g s_2 s_\beta^2-4 c_2 c_\beta s_\beta \lambda_T)\nonumber \\
\!\!\!\!\!&&+c_1 (g^\prime) (m_1^--m_1^+) s_1 (2 c_\beta g s_2 s_\beta+\sqrt{2} c_2 c_\beta^2 \lambda_T-\sqrt{2} c_2 s_\beta^2 \lambda_T)\bigg]\nonumber \\
\!\!\!\!\!R_{32}\!\!\!\!\!&=&\!\!\!\!\! \frac{c_W s_W M_Z^2 }{ g (g^\prime) (m_1^+-m_2^+) (m_2^+- m_1^-) \mu} \bigg[ c_1 (m_1^- - m_1^+) s_1 \lambda_S (\sqrt{2} c_\beta^2 g s_2 - \sqrt{2} g s_2 s_\beta^2 - 4 c_2 c_\beta s_\beta \lambda_T) \nonumber \\
&& +(m_2^+ - s_1^2 m_1^- - c_1^2 m_1^+)(-2 c_\beta g (g^\prime) s_2 s_\beta - \sqrt{2} c_2 c_\beta^2 (g^\prime) \lambda_T +    \sqrt{2} c_2 (g^\prime) s_\beta^2 \lambda_T) \bigg]\nonumber \\
\!\!\!\!\!R_{33}\!\!\!\!\!&=&\!\!\!\!\!c_2 -\frac{MZ^2 s_2 c_W^2  (2 c_2 c_\beta  s_2 s_\beta ((g)^2 + 2 \lambda_T^2) +\sqrt{2} g (c_2^2-s_2^2) (c_\beta^2-s_\beta^2)  \lambda_T) }{(g)^2 (m_2^--m_2^+) \mu} \nonumber \\
\!\!\!\!\!R_{34}\!\!\!\!\!&=&\!\!\!\!\!s_2 + \frac{MZ^2 c_2 c_W^2  (2 c_2 c_\beta  s_2 s_\beta ((g)^2 + 2 \lambda_T^2) +\sqrt{2} g (c_2^2-s_2^2) (c_\beta^2-s_\beta^2)  \lambda_T) }{(g)^2 (m_2^--m_2^+) \mu}\nonumber \\
\!\!\!\!\!R_{35}\!\!\!\!\!&=&\!\!\!\!\!-\frac{M_Z c_W}{g \mu} \bigg[ g s_2 s_\beta + \sqrt{2} c_2 c_\beta \lambda_T \bigg] \nonumber \\
\!\!\!\!\!R_{36}\!\!\!\!\!&=&\!\!\!\!\!\frac{M_Z c_W}{g \mu} \bigg[ g s_2 c_\beta - \sqrt{2} c_2 s_\beta \lambda_T \bigg] 
\end{eqnarray}
\begin{eqnarray}
\!\!\!\!\!R_{41}\!\!\!\!\!&=&\!\!\!\!\!\frac{c_W s_W M_Z^2 }{ g (g^\prime) (m_1^+-m_2^-) (m_2^--m_1^-) \mu} \bigg[ c_1 (g^\prime) (m_1^--m_1^+) s_1 (2 c_2 c_\beta g s_\beta+\sqrt{2} s_2 (-c_\beta^2+s_\beta^2) \lambda_T) \nonumber \\
\!\!\!\!\!&& + (c_1^2 m_1^- + s_1^2 m_1^+ - m_2^-  ) \lambda_S (\sqrt{2} c_2 g (c_\beta-s_\beta) (c_\beta+s_\beta)+4 c_\beta s_2 s_\beta \lambda_T) \bigg]\nonumber \\
\!\!\!\!\!R_{42}\!\!\!\!\!&=&\!\!\!\!\!\frac{c_W s_W M_Z^2 }{ g (g^\prime) (m_1^+-m_2^-) (m_2^--m_1^-) \mu} \bigg[ c_1 (m_1^--m_1^+) s_1 \lambda_S (\sqrt{2} c_2 g (c_\beta-s_\beta) (c_\beta+s_\beta)+4 c_\beta s_2 s_\beta \lambda_T) \nonumber \\
\!\!\!\!\!&&\!\!\!\!\!+(g^\prime) (c_1^2 m_1^+ + s_1^2 m_1^- -m_2^-) (2 c_2 c_\beta g s_\beta+\sqrt{2} s_2 (-c_\beta^2+s_\beta^2) \lambda_T)\bigg]\nonumber \\
\!\!\!\!\!R_{43}\!\!\!\!\!&=&\!\!\!\!\!-s_2 - \frac{MZ^2 c_2 c_W^2  (2 c_2 c_\beta  s_2 s_\beta ((g)^2 + 2 \lambda_T^2) +\sqrt{2} g (c_2^2-s_2^2) (c_\beta^2-s_\beta^2)  \lambda_T) }{(g)^2 (m_2^--m_2^+) \mu}\nonumber \\
\!\!\!\!\!R_{44}\!\!\!\!\!&=&\!\!\!\!\!c_2 - \frac{MZ^2 s_2 c_W^2  (2 c_2 c_\beta  s_2 s_\beta ((g)^2 + 2 \lambda_T^2) +\sqrt{2} g (c_2^2-s_2^2) (c_\beta^2-s_\beta^2)  \lambda_T) }{(g)^2 (m_2^--m_2^+) \mu}\nonumber \\
\!\!\!\!\!R_{45}\!\!\!\!\!&=&\!\!\!\!\!-\frac{M_Z c_W}{g' \mu} \bigg[ g c_2 s_\beta - \sqrt{2} s_2 c_\beta \lambda_T \bigg] \nonumber \\
\!\!\!\!\!R_{46}\!\!\!\!\!&=&\!\!\!\!\!\frac{M_Z c_W}{g' \mu} \bigg[ g c_2 c_\beta + \sqrt{2} s_2 s_\beta \lambda_T \bigg] 
\end{eqnarray}
\begin{eqnarray}
R_{51} &=& - \frac{M_Z s_W \lambda_S (c_\beta-s_\beta) }{g' \mu}\nonumber \\
R_{52} &=&  - \frac{M_Z s_W (c_\beta+s_\beta) }{ \sqrt{2}\mu}\nonumber \\
R_{53} &=& \frac{M_Z c_W \lambda_T (c_\beta-s_\beta) }{g \mu}\nonumber \\
R_{54} &=& \frac{M_Z c_W (c_\beta+s_\beta) }{ \sqrt{2}\mu}\nonumber \\
R_{55} &=& 1/\sqrt{2}\nonumber \\
R_{56} &=&  -1/\sqrt{2}
\end{eqnarray}
\begin{eqnarray}
R_{61} &=& - \frac{M_Z s_W \lambda_S (c_\beta+s_\beta) }{g' \mu}\nonumber \\
R_{62} &=& \frac{M_Z s_W (c_\beta-s_\beta) }{\sqrt{2} \mu}\nonumber \\
R_{63} &=&  \frac{M_Z c_W \lambda_T (c_\beta+s_\beta) }{g' \mu}\nonumber \\
R_{64} &=& -\frac{M_Z c_W (c_\beta-s_\beta) }{ \sqrt{2}\mu}\nonumber \\
R_{65} &=& 1/\sqrt{2}\nonumber \\
R_{66} &=& 1/\sqrt{2} 
\end{eqnarray}

The second order eigenvalues for the above states are given by
\begin{eqnarray}
m_{\psi_1} &=& m_1^+ - \frac{2 M_Z^2 s_W^2  \bigg[ c_\beta s_\beta ((g^\prime)^2 s_1^2  - 2 c_1^2 \lambda_S^2) +\sqrt{2} c_1 (g^\prime) s_1 (c_\beta^2-s_\beta^2) \lambda_S\bigg]}{(g')^2 \mu} \nonumber \\
m_{\psi_2} &=& m_1^- - \frac{2 M_Z^2 s_W^2  \bigg[ c_\beta s_\beta ((g^\prime)^2 s_1^2  - 2 c_1^2 \lambda_S^2)  -\sqrt{2} c_1 (g^\prime) s_1 (c_\beta^2-s_\beta^2) \lambda_S  \bigg]}{(g')^2 \mu} \nonumber \\
m_{\psi_3} &=& m_2^+ - \frac{2 M_Z^2 c_W^2  \bigg[ c_\beta s_\beta ( c_2^2  g^2  - 2  s_2^2  \lambda_T^2 ) +\sqrt{2} c_2 g s_2 (c_\beta^2-s_\beta^2) \lambda_T \bigg]}{g \mu} \nonumber \\
m_{\psi_4} &=& m_2^- - \frac{2 M_Z^2 c_W^2  \bigg[ c_\beta s_\beta ( c_2^2  g^2  - 2  s_2^2  \lambda_T^2 ) -\sqrt{2} c_2 g s_2 (c_\beta^2-s_\beta^2) \lambda_T \bigg]}{g \mu} \nonumber \\
m_{\psi_5} &=& \mu + \frac{2 M_Z^2 s_W^2}{4\mu (g^\prime)^2} \bigg[(g^\prime)^2 + 2 (c_\beta - s_\beta)^2 \lambda_S^2\bigg] + \frac{2 M_Z^2 c_W^2}{4\mu g^2}\bigg[g^2  + 2 (c_\beta - s_\beta)^2 \lambda_T^2 \bigg] \nonumber \\
m_{\psi_6} &=& -\mu - \frac{2 M_Z^2 s_W^2}{4\mu (g^\prime)^2} \bigg[(g^\prime)^2 (c_\beta - s_\beta)^2 + 2  \lambda_S^2 \bigg] - \frac{2 M_Z^2 c_W^2}{4\mu g^2} \bigg[ g^2 (c_\beta - s_\beta)^2 + 2  \lambda_T^2\bigg]
\end{eqnarray}

\section{Rotation Matrices and Eigenvalues for Mostly Higgsino Neutralinos}
\label{Appendix:Higgsino}

In this section we give the rotation matrix to first order for the case that $\mu$ is much less than the other masses. 
\begin{eqnarray}
R_{11} &=& c_1 \nonumber \\
R_{12} &=& s_1 \nonumber \\
R_{13} &=& 0 \nonumber \\
R_{14} &=& 0 \nonumber \\
R_{15} &=& \frac{M_Z s_W}{g^\prime m_1^+} (\sqrt{2} \lambda_S c_1 s_\beta-g^\prime c_\beta s_1)
\nonumber \\ 
R_{16} &=& \frac{M_Z s_W}{g^\prime m_1^+} (\sqrt{2} \lambda_S c_\beta c_1+g^\prime s_\beta s_1) 
\end{eqnarray}
\begin{eqnarray}
R_{21} &=& -s_1 \nonumber \\
R_{22} &=& c_1 \nonumber \\
R_{23} &=& 0 \nonumber \\
R_{24} &=& 0 \nonumber \\
R_{25} &=& -\frac{M_Z s_W}{g^\prime m_1^-} (g^\prime c_\beta c_1+\sqrt{2} \lambda_S s_\beta s_1) \nonumber \\ 
R_{26} &=& \frac{M_Z s_W}{g^\prime m_1^-} (g^\prime c_1 s_\beta-\sqrt{2} \lambda_S c_\beta s_1)
\end{eqnarray}
\begin{eqnarray}
R_{31} &=& 0 \nonumber \\
R_{32} &=& 0 \nonumber \\
R_{33} &=& c_2 \nonumber \\
R_{34} &=& s_2 \nonumber \\
R_{35} &=& \frac{M_Z c_W}{g m_2^+} (-\sqrt{2} \lambda_T c_2 s_\beta+g c_\beta s_2)
 \nonumber \\ 
R_{36} &=& -\frac{M_Z c_W}{g m_2^+} (\sqrt{2} \lambda_T c_\beta c_2+g s_\beta s_2)
\end{eqnarray}
\begin{eqnarray}
R_{41} &=& 0 \nonumber \\
R_{42} &=& 0 \nonumber \\
R_{43} &=& -s_2 \nonumber \\
R_{44} &=& c_2 \nonumber \\
R_{45} &=& \frac{M_Z c_W}{g m_2^-} (g c_\beta c_2+\sqrt{2} \lambda_T s_\beta s_2) \nonumber \\ 
R_{46} &=& -\frac{M_Z c_W}{g m_2^-} (-g c_2 s_\beta+\sqrt{2} \lambda_T c_\beta s_2)
\end{eqnarray}
\begin{eqnarray}
R_{51} &=& \frac{M_Z s_W}{2 g' m_1^- m_1^+} \bigg[2 \lambda_S ( m_1^-  c_1^2 +  m_1^+ s_1^2) (c_\beta-s_\beta)+\sqrt{2} (g^\prime) (m_1^--m_1^+) c_1 (c_\beta+s_\beta) s_1 \bigg] \nonumber \\
R_{52} &=& \frac{M_Z s_W}{2 g' m_1^- m_1^+} \bigg[\sqrt{2} (g^\prime) (c_\beta+s_\beta) (m_1^+ c_1^2+m_1^- s_1^2)+(m_1^--m_1^+) \lambda_S (c_\beta-s_\beta) 2 c_1 s_1\bigg] \nonumber \\
R_{53} &=& -\frac{M_Z c_W}{2 g' m_2^- m_2^+} \bigg[ 2\lambda_T (c_\beta-s_\beta) (c_2^2 m_2^-+ s_2^2 m_2^+) + \sqrt{2} g (m_2^--m_2^+) (c_\beta+s_\beta) c_2 s_2 \bigg]\nonumber \\
R_{54} &=& -\frac{M_Z c_W}{2 g' m_2^- m_2^+} \bigg[ \sqrt{2} g ( s_2^2 m_2^- + c_2^2 m_2^+) (c_\beta+s_\beta)+  2\lambda_T(m_2^--m_2^+)  (c_\beta-s_\beta) c_2 s_2 \bigg] \nonumber \\
R_{55} &=& \frac{1}{\sqrt{2}} + R_H \nonumber \\
R_{56} &=& -\frac{1}{\sqrt{2}} + R_H 
\end{eqnarray}
\begin{eqnarray}
R_{61} &=& \frac{M_Z s_W}{2 g' m_1^- m_1^+} \bigg[ -2(m_1^- c_1^2 +m_1^+ s_1^2 ) \lambda_S (c_\beta+s_\beta) +\sqrt{2} (g^\prime) (m_1^--m_1^+) c_1 (c_\beta-s_\beta) s_1
\bigg] \nonumber \\
R_{62} &=& \frac{M_Z s_W}{2 g' m_1^- m_1^+} \bigg[ -2 (m_1^--m_1^+) \lambda_S c_1 (c_\beta+s_\beta) s_1+\sqrt{2} (g^\prime) (c_\beta-s_\beta) (m_1^+ c_1^2+m_1^- s_1^2)
\bigg] \nonumber \\
R_{63} &=& \frac{M_Z c_W}{2 g' m_2^- m_2^+} \bigg[  2( c_2^2m_2^-+ s_2^2 m_2^+) \lambda_T (c_\beta+s_\beta)-\sqrt{2} g (m_2^--m_2^+) (c_\beta-s_\beta) 2 c_2 s_2) \bigg]\nonumber \\
R_{64} &=& \frac{M_Z c_W}{2 g' m_2^- m_2^+} \bigg[-2 \sqrt{2} g ( s_2^2 m_2^- + c_2^2 m_2^+) (c_\beta-s_\beta)+2 (m_2^--m_2^+) \lambda_T (c_\beta+s_\beta) 2 c_2 s_2\bigg] \nonumber \\
R_{65} &=& \frac{1}{\sqrt{2}} - R_H \nonumber \\
R_{66} &=& \frac{1}{\sqrt{2}} - R_H
\end{eqnarray}
where
\begin{eqnarray}
R_H &=& \frac{M_Z^2}{8 \sqrt{2} g^2 (g^\prime)^2 m_1^- m_1^+ m_2^- m_2^+ \mu} \times \bigg[ \\
&& c_{2\beta} \bigg\{(g^\prime)^2 m_1^- m_1^+ (-(m_2^-+m_2^+) (g^2-2 \lambda_T^2) 
 +(m_2^--m_2^+) (g^2+2 \lambda_T^2) (c_2^2- s_2^2)) c_W^2 \nonumber \\ &&+g^2 m_2^- m_2^+ (-(m_1^-+m_1^+) ((g^\prime)^2-2 \lambda_S^2)+(m_1^--m_1^+) ((g^\prime)^2+2 \lambda_S^2) (c_1^2 -  s_1^2)) s_W^2\bigg\} \nonumber \\
&&+s_{2\beta}g (g^\prime) 2 \sqrt{2}  \bigg\{  (g^\prime) m_1^- m_1^+ (m_2^--m_2^+) \lambda_T c_W^2 2 c_2 s_2+g (m_1^--m_1^+) m_2^- m_2^+ \lambda_S 2 c_1 s_1 s_W^2\bigg\}\bigg] \nonumber 
\end{eqnarray}
The non-$LSP$ eigenvalues are
\begin{eqnarray}
m_{\psi_1} &=& m_1^+ + \frac{2 M_Z^2 s_W^2  \bigg[ 2 \lambda_S^2 c_1^2+(g^\prime)^2 s_1^2 \bigg]}{(g')^2 \mu} \nonumber \\
m_{\psi_2} &=& m_1^- + \frac{2 M_Z^2 s_W^2  \bigg[ c_1^2 (g')^2 + 2 s_1^2 \lambda_S^2  \bigg]}{(g')^2 \mu} \nonumber \\
m_{\psi_3} &=& m_2^+ + \frac{2 M_Z^2 c_W^2  \bigg[ s_2^2 g^2 + 2 c_2^2 \lambda_T^2 \bigg]}{g \mu} \nonumber \\
m_{\psi_4} &=& m_2^- + \frac{2 M_Z^2 c_W^2  \bigg[ c_2^2 g^2 + 2 s_2^2 \lambda_T^2  \bigg]}{g \mu} \nonumber \\
\end{eqnarray}
while the $LSP$ and $NLSP$ are given by
\begin{eqnarray}
m_{\psi_5} &=& \mu  -\frac{M_Z^2  c_W^2 (\sqrt{2} g c_2 (c_\beta+s_\beta)+2 \lambda_T (-c_\beta+s_\beta) s_2)^2}{4g^2 m_2^-} \nonumber \\
&&-\frac{M_Z^2  c_W^2 (2 \lambda_T c_2 (c_\beta-s_\beta)+\sqrt{2} g (c_\beta+s_\beta)s_2)^2}{4g^2 m_2^+} \nonumber \\
&&-\frac{M_Z^2  s_W^2 (\sqrt{2} (g^\prime) c_1 (c_\beta+s_\beta)+2 \lambda_S (-c_\beta+s_\beta) s_1)^2}{4(g^\prime)^2 m_1^-}\nonumber \\
&&-\frac{M_Z^2 s_W^2 (2 \lambda_S c_1 (c_\beta-s_\beta)+\sqrt{2} (g^\prime) (c_\beta+s_\beta) s_1)^2 }{4(g^\prime)^2 m_1^+}\nonumber \\
m_{\psi_6} &=& -\mu -\frac{M_Z^2  c_W^2 (\sqrt{2} g c_2 (c_\beta-s_\beta)+2 \lambda_T (c_\beta+s_\beta)s_2)^2}{4g^2 m_2^-}\nonumber \\ 
&&-\frac{M_Z^2  c_W^2 (-2 \lambda_T c_2 (c_\beta+s_\beta)+\sqrt{2} g (c_\beta-s_\beta) s_2)^2}{4g^2 m_2^+}\nonumber \\
&&-\frac{M_Z^2 s_W^2 (\sqrt{2} (g^\prime) c_1 (c_\beta-s_\beta)+2 \lambda_S (c_\beta+s_\beta) s_1)^2 }{4(g^\prime)^2 m_1^-}\nonumber \\
&&-\frac{M_Z^2 s_W^2 (2 \lambda_S c_1 (c_\beta+s_\beta)+\sqrt{2} (g^\prime) (-c_\beta+s_\beta) s_1)^2 }{4(g^\prime)^2 m_1^+}
\end{eqnarray}

\section{Couplings of the LSP}

The neutralino-chargino-W interactions depend on the
$\wino,\winop,\tilde{h}$ components of the neutralino,
\begin{equation}
{\cal L}=\overline{\tilde\chi^-}_a \gamma^\mu\left(C^{ai}_L (1-\gamma_5)
+C^{ai}_R (1+\gamma_5)\right)
 \overline{\tilde{\chi}^0_i}W_\mu^- +h.c.
\end{equation}
where
\begin{eqnarray}
C^{ai}_L=\frac{e}{4 s_W}\left(2 N_{i4} U_{a2} +2 N_{i3}
U_{a1}+\sqrt{2}N_{i5}U_{a3}\right)\nonumber\\
C^{ai}_R=\frac{e}{4 s_W}\left(2 N_{i4} V_{a2} +2 N_{i3}
V_{a1}-\sqrt{2}N_{i6}V_{a3}\right)
\end{eqnarray}
When only gaugino Dirac mass are presents and
when $m_{1D},m_{2D}<<\mu$ the dominant contribution comes from the 
$N_{13} U_{11}$ term.

The Majorana neutralino coupling to the Z is driven by the higgsino
component, as in the MSSM,
\begin{equation}
{\cal L}=\frac{1}{2}\overline{\tilde{\chi}^0}_i \gamma^\mu\gamma_5 C^{ij}_Z
 \overline{\tilde{\chi}^0}_i Z_\mu +h.c.
\end{equation}
\begin{equation}
C^{ii}_Z=\frac{g}{2c_W} \left(|N_{i6}|^2-|N_{i5}|^2\right)
\end{equation}

The neutralino couplings to the light Higgs also depend on  the
$\wino,\winop,\tilde{h}$ components of the neutralino
\begin{equation}
{\cal L}=\frac{1}{2}\overline{\tilde{\chi}^0}_i  C^{ij}_h
 \overline{\tilde{\chi}^0}_i h +h.c.
\end{equation}
where
\begin{eqnarray}
C^{ii}_h&=&\frac{-1}{ s_W c_W} \left[  e\left( c_W N_{i4}-s_W
N_{i2}\right) (Z_{h_{11}}N_{i5}- Z_{h_{12}} N_{i6})
 \right.\nonumber\\
&&+\left. \sqrt{2}s_W c_W 
\left(N_{i1}\lambda_S-N_{i3}\lambda_T)\right)\left(Z_{h_{12}} N_{i5}+ 
Z_{h_{11}} N_{i6}
\right) \right]
\end{eqnarray}
Here $Z_{h}$ is the scalar mixing matrix.

\end{document}